\documentclass[fleqn,usenatbib,useAMS]{mnras}

\usepackage{subfiles}
\usepackage{amsmath}	% Advanced maths commands
\usepackage{amssymb}	% Extra maths symbols
\usepackage{multicol}        % Multi-column entries in tables
\usepackage{bm}		% Bold maths symbols, including upright Greek
\usepackage{pdflscape}	% Landscape pages
\usepackage{microtype}      % microtypography
\usepackage{float}
\usepackage{graphicx}
\usepackage{caption}
\usepackage{natbib}
\usepackage{lipsum}
\usepackage{calc}  
\usepackage{enumitem}  
\usepackage[T1]{fontenc}
\usepackage{ae,aecompl}
\usepackage{newtxtext,newtxmath}
\usepackage{stackengine}
\usepackage{booktabs,makecell}
\usepackage{siunitx}
\usepackage{pifont}
\usepackage{tabularx}
\usepackage{cleveref}
\usepackage{multirow}
\usepackage{boldline}
\usepackage{array}
\usepackage{xfrac}
\usepackage{subcaption}
\usepackage{orcidlink}
\usepackage{adjustbox}

\usepackage{xcolor}
\defcitealias{Etherington2022}{E22}
\defcitealias{Cao2022}{C22}

\title[`External shear' is not shear]{Strong gravitational lensing's `external shear' is not shear}

\author[Etherington et al.]{\parbox{\textwidth}{Amy Etherington$^{1,2}$\orcidlink{0000-0003-3185-6592}, 
James W.\ Nightingale$^{1,2}$\thanks{Contact e-mail: \href{mailto:james.w.nightingale@durham.ac.uk}{james.w.nightingale@durham.ac.uk}}\orcidlink{0000-0002-8987-7401},
Richard Massey$^{1,2}$\orcidlink{0000-0002-6085-3780},
Sut-Ieng Tam$^{3}$\orcidlink{0000-0002-6724-833X}, 
XiaoYue Cao$^{4,5}$\orcidlink{0000-0003-4988-9296},
Anna Niemiec$^{1}$, 
Qiuhan He$^{2}$\orcidlink{0000-0003-3672-9365}, 
Andrew Robertson$^{6}$\orcidlink{0000-0002-0086-0524}, 
Ran Li$^{5,4}$,  
Aristeidis Amvrosiadis$^{2}$, 
Shaun Cole$^{2}$,
Jose M.\ Diego$^{7}$,
Carlos S.\ Frenk$^{2}$,
Brenda L.\ Frye$^{8}$,
David Harvey$^{9}$,
Mathilde Jauzac$^{1,2,10,11}$,
Anton M.\ Koekemoer$^{10}$\orcidlink{0000-0002-6610-2048},
David J.\ Lagattuta$^{1}$,
Marceau Limousin$^{11}$, 
Guillaume Mahler$^{1}$, 
Ellen Sirks$^{12}$\orcidlink{0000-0002-7542-0355} \&
Charles L.\ Steinhardt$^{13}$
}\\
\\
$^{1}$Department of Physics, Centre for Extragalactic Astronomy, Durham University, South Rd, Durham, DH1 3LE, UK \\
$^{2}$Department of Physics, Institute for Computational Cosmology, Durham University, South Road, Durham DH1 3LE, UK \\
$^{3}$Academia Sinica Institute of Astronomy and Astrophysics (ASIAA), No.\ 1, Sec.\ 4, Roosevelt Road, Taipei 10617, Taiwan \\
$^{4}$School of Astronomy and Space Science, University of Chinese Academy of Sciences, Beijing 100049, China \\
$^{5}$National Astronomical Observatories, Chinese Academy of Sciences, 20A Datun Road, Chaoyang District, Beijing 100012, China\\
$^{6}$Jet Propulsion Laboratory, California Institute of Technology, 4800 Oak Grove Drive, Pasadena, CA 91109, USA\\
$^{7}$Instituto de Física de Cantabria (CSIC-UC), Avda. Los Castros s/n, 39005 Santander, Spain\\
$^{8}$Department of Astronomy/Steward Observatory, University of Arizona, 933 N Cherry Ave., Tucson, AZ 85721, USA\\
$^{9}$Laboratoire d’Astrophysique, EPFL, Observatoire de Sauverny, 1290 Versoix, Switzerland\\
$^{10}$Space Telescope Science Institute, 3700 San Martin Dr., Baltimore, MD 21218, USA\\
$^{11}$Aix Marseille Univ, CNRS, CNES, LAM, F-13388 Marseille, France\\
$^{12}$Sydney Consortium for Particle Physics and Cosmology, School of Physics, The University of Sydney, NSW 2006, Australia\\
$^{13}$Niels Bohr Institute, University of Copenhagen, Lyngbyvej 2, K\o benhavn \O\ 2100, Denmark
}

\date{}

\pubyear{2023}

\begin{document}
\label{firstpage}
\pagerange{\pageref{firstpage}--\pageref{lastpage}}
\maketitle

\begin{abstract}
The distribution of mass in galaxy-scale strong gravitational lenses is often modelled as an elliptical power law plus `external shear', which notionally accounts for neighbouring galaxies and cosmic shear. We show that it does not. 
Except in a handful of rare systems, the best-fit values of external shear do not correlate with independent measurements of shear: from weak lensing in 45 Hubble Space Telescope images, or in 50 mock images of lenses with complex distributions of mass. Instead, the best-fit shear is aligned with the major or minor axis of 88\% of lens galaxies; and the amplitude of the external shear increases if that galaxy is disky.
We conclude that `external shear' attached to a power law model is not physically meaningful, but a fudge to compensate for lack of model complexity.
Since it biases other model parameters that {\it are} interpreted as physically meaningful in several science analyses (e.g.\ measuring galaxy evolution, dark matter physics or cosmological parameters), we recommend that future studies of galaxy-scale strong lensing should employ more flexible mass models.

%Galaxy-scale strong gravitational lenses are commonly modelled using elliptical mass distributions with additional `external shear' parameters that are justified as accounting for perturbations such as neighbouring galaxies and cosmic shear. We show that this assumption is oversimplified. For a sample of \rjm{39??} SLACS lenses the external shear measured by strong lensing does not correlate with the true shear along that line of sight, as measured independently from weak lensing. The strong lensing external shears are systematically overestimated by a magnitude of 0.06 on average. We demonstrate that these shears are in fact compensating for the lack of azimuthal degrees of freedom in the power law elliptical mass distribution that we assume for the lens galaxy. We do this by comparing the behaviour of the shears to those incorrectly inferred in a mock sample of data that were simulated without external shear. A number of results suggest that the complexity of the observed galaxies is greater than the simulated MGE+NFW mock sample, including the lack of a consistently recovered position angle when we fit models without the external shear parameters, and measurements of significant $b_4$ multipole components in the critical curves. We suspect this implies twists are present in the mass distribution. Future strong lensing studies will require more complex mass models to appropriately describe the lens galaxy, and avoid biases on high precision measurements of galaxy masses, cosmological parameters $H_0$, and dark matter substructures

\end{abstract}


\section{Introduction}\label{Introduction}

Gravitational lensing is the deflection of light rays by nearby concentrations of matter and their associated gravitational fields. If the light ray should pass straight through an object as massive as a galaxy, it can be deflected along multiple routes around the galaxy, and appear distorted into arcs or an `Einstein ring'. Such galaxy-scale strong lensing has been used to infer the distribution of mass in massive elliptical galaxies \citep{Gavazzi2007, Koopmans2009, Auger2010, Sonnenfeld2013b, Bolton2012, Etherington2022a}, to infer their dark matter content, stellar mass-to-light ratios, and inner structure \citep{Massey2010, Sonnenfeld2012, Oldham2018, Nightingale2019, Shu2015, Shu2016c}. If the background source is variable, measurements of time delays between multiple images can be used to measure cosmological parameters \citep{Kilbinger2015} or the Hubble constant \citep{Suyu2017, Wong2019, Harvey2020b, Gomer2021}. If the lens galaxy contains small substructures, which would be a smoking gun of the `clumpy' Cold Dark Matter (CDM) model, they would also perturb the multiple images \citep{Natarajan2004, Vegetti2010, Vegetti2014b, Li2016, Li2017, Hezaveh2016, Ritondale2019, Despali2019, HeAmy2022a, HeAmy2022b, Amorisco2021, Nightingale2022}. 

When modelling the distribution of mass to fit strong lensing data, two additional free parameters are frequently included. The (amplitude and angle of) `external shear' is intended to represent the cumulative deflection of light by all other gravitational potentials along the line of sight. Indeed, the best-fit value of external shear matched a model of the line of sight for three of six galaxy lenses studied by \citet{Wong2011}. Subsequent papers have even proposed using external shear as relatively high signal-to-noise measurements of the `cosmic' shear along individual lines of sight \citep{Birrer2017, Desprez2018, Kuhn2020, Fleury2021, Hogg2022}. However, the best-fit values of external shear did not match the lines of sight to the three other galaxies studied by \citet{Wong2011} --- and, in general, best-fit values are much larger than expected for both galaxy lenses \citep{Keeton1997, Witt1997, Hilbert2007, Suyu2010, Barrera2021} and cluster lenses \citep{Robertson2020b,Limousin2022}. Using mock observations, \citet[][hereafter C22]{Cao2022} demonstrated that the best-fit shear can be incorrect if the model of the mass distribution is missing complexity.
%There is evidence that this shear assumption can also impact estimates of the Hubble constant \citep{Gomer2021}.

In this paper we compare the external shear measured from Hubble Space Telescope (HST) imaging --- of strong lensing galaxies from the SLACS survey \citep{Bolton2008a}, GALLERY survey \citep{Shu2016b}, and four lenses in clusters --- against independent measurements of the shear along the same line-of-sight, observed as weak lensing of adjacent galaxies. 
To gain understanding, we also analyse \citetalias{Cao2022}'s mock HST imaging, generated without external shear, but fitted with shear as a free parameter. 

Comparing independent measurements of shear will test the hypothesis that strong lensing external shear is a real, physical quantity. 
Strong and weak lensing measurements average over different spatial scales and are obtained at different redshifts, so they might not be identical; but they should be strongly correlated. Analysing three lenses in the COSMOS field \citep{Faure2008} with sophisticated statistics, \cite{Kuhn2020} measured a smaller covariance between strong and weak lensing shears than the difference between individual systems, indicating that more data were required. In this work, with a much larger sample of galaxies, we simply aim to detect a correlation between the two probes.

%To gain further understanding of the external shears measured from real observations, we also compare to fits of mock data by \citetalias{Cao2022}. These mock data were generated without external shear, but including it improved the fit of a power law + external shear (PL + ext) mass model. 

We define relevant concepts from lensing theory in Section~\ref{Theory}. We then describe our observed and mock data in Section~\ref{Data} and our analysis methods in Section~\ref{Method}. We present our results in Section~\ref{Results} and investigate possible causes in Section~\ref{section: analysis}. We interpret these in a wider context and conclude in Section~\ref{Discussion}.  Throughout this paper, we assume a Planck 2015 cosmological model \citep{Ade2016}.

\section{Gravitational lensing theory}\label{Theory}
Gravitational lensing describes the deflection of light rays from distant sources, by matter along its path to an observer, through angle $\bm{\alpha}$. This maps 2D coordinates of light on the distant source plane $\bm{\beta}$ to coordinates where they are observed on the foreground image plane $\bm{\theta}=(\theta_1,\theta_2)$, via the lens equation 
\begin{equation}\label{eq: lens equation}
    \bm{\beta} = \bm{\theta}-\bm{\alpha}(\bm{\theta}).
\end{equation}

If the gravitational lens is much thinner than its angular diameter distance from the observer $D_{\rm l}$, its distribution of mass can be treated as a 2D surface density projected along the line-of-sight $\Sigma(\bm{\theta})= %\Sigma(x/D_{\rm l}, D_{y/\rm l})=
\int\rho(D_{\rm l}\theta_1,D_{\rm l}\theta_2,z)\,dz$. Following the notation of \cite{NarayanBartelmann1996}, the deflection angle $\bm{\alpha}=\bm{\nabla}\psi$ is the vector gradient of a 2D lensing potential 
\begin{align}\label{eq:phi(Sigma)}
    \psi(\bm{\theta}) &\equiv 2 \frac{D_{\rm ls}}{D_{\rm l} D_{\rm s}} \frac{2G}{c^2} \int \Sigma(\bm{\theta}') \ln{|\bm{\theta}-\bm{\theta}'|}~d^2\bm{\theta}' \\
    ~ &= 2 \frac{D_{\rm ls}}{D_{\rm l} D_{\rm s}} \int \Phi(D_{\rm l}\bm\theta, z)~dz~,
\end{align}
where $\Phi$ is the 3D Newtonian potential, and the prefactor (which involves angular diameter distances to the lens, to the source, and from the lens to the source) reflects the geometrical efficiency of a lens: peaking if it is half-way between the source and the observer.

The Jacobian of the lens equation~\eqref{eq: lens equation} is thus
\begin{equation}\label{eq: Jacobian}
    A_{ij}\equiv\frac{\partial \beta_i}{\partial\theta_j} = \delta_{ij} - \frac{\partial\alpha_{i}}{\partial\theta_j} =\delta_{ij} - \frac{\partial^2\psi}{\partial \theta_i \theta_j}.
\end{equation}

\subsection{Weak lensing}\label{weak lensing theory}

%If the gravitational field is constant on scales larger than the source, the Jacobian for this (small) source galaxy can be treated as constant
If the source is much smaller than the scale of local variations in the gravitational field, the Jacobian can be approximated as constant
\begin{align}\label{eq: linear Jacobian}
    A_{ij} &\approx 
    \begin{pmatrix}
    1-\kappa-\gamma_1 & -\gamma_2 \\
    -\gamma_2 & 1-\kappa+\gamma_1
    \end{pmatrix}\\
    &= (1-\kappa)\begin{pmatrix}
    1 & 0 \\
    0 & 1
    \end{pmatrix}
    -\gamma\begin{pmatrix}
    \mathrm{cos}2\phi & \mathrm{sin}2\phi \\
    \mathrm{sin}2\phi & \mathrm{cos}2\phi
    \end{pmatrix},
\end{align}
where the convergence 
\begin{equation}
    \kappa=\frac{1}{2}\bm{\nabla}^2\psi ~,%\ll 1~,
\end{equation} 
and where the two components of shear 
\begin{equation}
    \Big(\gamma_1, ~ \gamma_2\Big) = 
    \left( \frac{1}{2} \left( \frac{\partial^2\psi}{\partial \theta_1^2} - \frac{\partial^2\psi}{\partial \theta_2^2} \right) ,  ~ \frac{\partial^2\psi}{\partial \theta_1 \partial \theta_2} \right)
\end{equation}
can also be expressed in terms of shear amplitude $\gamma^2=\gamma_1^2+\gamma_2^2$ and angle $\phi=\,$\textonehalf$\,\arctan{(\gamma_2/\gamma_1)}$. 
The convergence magnifies a source, and the shear changes its shape.
Strictly, %due to the invariance of the Jacobian matrix under the transformation $\bm{A}\rightarrow\lambda\bm{A}$ \citep[known as the mass sheet degeneracy;][]{Schneider2013}, 
these quantities are only observable in combination as `reduced shear' $g_i = {\gamma_i}/{(1-\kappa)}$. However, in the weak lensing regime, $\kappa\ll 1$, so $g_i\approx\gamma_i$.

\if{false}
Since the unlensed shape of a background galaxy $\epsilon^\mathrm{int}$ is unknown (and not circular), the measured ellipticity observed on the sky is
\begin{equation}\label{eq: WL ellipticity}
    \epsilon_i = \epsilon_i^\mathrm{int} + Gg_i,
\end{equation}
where scalar $G$ is the shear `responsivity' \citep{Rhodes2000}.

It is possible to measure the mean reduced shear in a patch of sky, by averaging galaxies' apparent shapes $\epsilon$, under the assumption that their intrinisic shapes are randomly oriented, i.e.\ $\langle\epsilon_i^\mathrm{int}\rangle\approx0$, so
\begin{equation}\label{eq: WL shear estimator}
    \langle g_i\rangle 
    = \left\langle\frac{\epsilon_i - \epsilon_i^\mathrm{int}}{G}\right\rangle
    \approx \left\langle\frac{\epsilon_i}{G}\right\rangle - \left\langle\frac{\epsilon_i^\mathrm{int}}{G}\right\rangle
    \approx \frac{\langle\epsilon_i\rangle}{\langle G \rangle}.
\end{equation}
In the weak lensing limit $\kappa\ll 1$, the reduced shear approximates the shear itself $g\approx\gamma$. Although the line-of-sight directly through the galaxy-scale lenses we are interested in is \textit{not} in the weak lensing regime, we assume this approximation still holds since the vast majority of adjacent lines of sights will not be strongly lensed. Thus, the weak lensing shear measurements are statistical in nature, limited by the randomness in the distribution of the unknown intrinsic ellipticites termed `shape noise' 
\begin{equation}
    \sigma_\mathrm{int}^2=\left\langle\left(\frac{\epsilon_i^\mathrm{int}}{G}\right)^2\right\rangle.
\end{equation}
\fi

\subsection{Strong lensing}\label{strong lensing theory}

If light from one side of a source is deflected differently to light from the other side, it can appear distorted in the image plane as an arc; it is also possible to see multiple images of a single source, if more than one solution exists with different $\bm{\alpha}$ and $\bm{\theta}$. To reconstruct $\bm{\alpha}(\bm{\theta})$ it is usual to note that 
\begin{align}
    \kappa(\bm\theta) &= \frac{\Sigma(\bm\theta)}{\Sigma_\mathrm{crit}} & 
    \mathrm{with~constant}&  &\Sigma_\mathrm{crit}=&\frac{{\rm c}^2}{4{\rm \pi} {\rm G}}\frac{D_{\rm s}}{D_{\rm l} D_{\rm ls}}\,,
    \label{eq: sigma crit}
\end{align}
which is equal to the mean surface mass density within the Einstein radius, $R_\mathrm{Ein}$. For axisymmetric lenses the Einstein radius is uniquely defined by the radius of the circular tangential critical curve that is produced where the magnification diverges in the lens plane. This occurs where the tangential eigenvalue of the Jacobian (equation~\ref{eq: Jacobian}) 
$\lambda_\mathrm{t}=1-\kappa-\gamma$ is equal to zero. For asymmetric lenses, the definition of Einstein radius must be generalised; we choose to use the \textit{effective} Einstein radius
\begin{equation}
    R_\mathrm{Ein,eff}=\sqrt{\frac{A}{\mathrm{\pi}}},
    \label{eq: einstein rad}
\end{equation}
where $A$ is the area enclosed by the tangential critical curve.

When considering (typically Early-type) galaxy-scale lenses, it is common practise to parameterise the surface mass distribution as an elliptical power law \citep{Suyu2012}
\begin{equation}\label{eq: power law}
    \Sigma(\theta_1,\theta_2; b, q, \gamma) =
    \frac{3-\gamma}{1+q}\left(\frac{b}{\sqrt{\theta_1^2+\theta_2^2/q^2}}\right)^{\gamma-1},
\end{equation} where $b\geq0$ is the angular scale length (referred to in some papers as the Einstein radius, but distinct from the more robust \textit{effective} Einstein radius in Equation~\ref{eq: einstein rad}), $0 < q \leq 1$ is the projected minor to major axis ratio of the elliptical isodensity contours, and (confusingly denoted) $\gamma$ is the logarithmic slope of the mass distribution in 3D (for an `isothermal' distribution, $\gamma=2$). If we also allow the mass to be translated to central coordinates $(\theta_{1}^{\rm c}, \theta_{2}^{\rm c})$ and reoriented to position angle $\phi^\mathrm{PL}$, which we measure counterclockwise from the positive $\theta_1$-axis, the model has six free parameters.

The primary lens may not be the only source of shear. Any `external' component due to other galaxies or clusters near the lens or along the ray path, and constant on scales larger than $b$ (rather than the size of the source)  is modelled as two more free parameters
\begin{equation}
\Big(\gamma^{\mathrm{ext}}_1, ~ \gamma^{\mathrm{ext}}_2\Big) = 
\gamma^\mathrm{ext} \Big( \cos{(2\phi^\mathrm{ext})}, ~ \sin{(2\phi^\mathrm{ext})} \Big)~,  
\label{eq: shear strong}
\end{equation}
where $\gamma^\mathrm{ext}$ is the amplitude and $\phi^\mathrm{ext}$ is the angle of the shear measured counterclockwise from the $\theta_1$-axis. This is applied as an additional component of $\bm{\alpha}(\bm{\theta})$. It does not change $\kappa(\bm{\theta})$.
%This is applied as a coordinate transformation exactly as equation~\eqref{eq: linear Jacobian}, but corresponds to shear that is constant on scales larger than $b$, rather than the source. 

%\subsection{Comparing different shear measurements}
%Note that strongly lensed and weakly lensed sources may be at different redshifts. However, for a fixed lens (mass distribution and redshift), $\gamma_i\propto D_{\rm ls}/D_{\rm s}$, so %the quantity $\gamma_i D_{\rm s}/D_{\rm ls}$
%\begin{equation}
%\gamma_i^\infty \equiv \frac{D_{\rm s}}{D_{\rm ls}}\,\gamma_i 
%\end{equation}
%(i.e.\ the shear experienced by a source at infinite distance) should be constant, regardless of how it is measured.
   
\section{Data}\label{Data} 
\subsection{Mock lens galaxies}\label{literature}
We analyse a set of 50 mock lens images, representative of data from the SLACS survey. They were generated by \citetalias{Cao2022} for an investigation into the systematic errors induced by the elliptical power-law model. We summarise the simulation procedure below; a detailed description can be found in Section~2.4 of that paper.

The surface mass density of the lens galaxy comprises two components: a dark matter halo, parameterised by the spherical generalised Navarro, Frenk \& White (gNFW) profile \citep{Navarro1997,Zhao1996,Cappellari2013}, plus visible stellar matter, parameterised by a Multiple Gaussian Expansion \citep[MGE;][]{Cappellari2002}. The model parameters of the gNFW and MGE profiles of each lens galaxy are set to the best-fit parameters from fits of these distributions to SDSS-MaNGA stellar dynamics data, derived by \cite{Li2019} using the Jeans anisotropic model (JAM) method. The position angle of each Gaussian component in the MGE is fixed, however their axis ratios are free to vary, allowing for elliptical gradients in the mass distribution. 

The light distribution of the source galaxy is modelled by a single S\'ersic profile \citep{Graham2005} with effective radius $R_\mathrm{eff}=0.15\arcsec$, S\'ersic index $n=1$, and axis ratio $q=0.7$. The position in the source plane $(x_s, y_s)$ is drawn from a Gaussian distribution with mean 0$\arcsec$ and standard deviation 0.1$\arcsec$, and the position angle is uniformly selected between $0^\circ-180^\circ$. The light from the source galaxy is ray-traced from the source plane at $z=0.6$ to the image plane through the lens equation (equation~\ref{eq: lens equation}), to simulate its lensed appearance. Further, to mimic observational effects, the image is convolved with a Gaussian PSF with 0.05$\arcsec$ standard deviation, and sampled by 0.05$\arcsec$ square pixels. A flat background sky of 0.1 electrons per second is assumed, and an exposure time of 840 seconds is used to add Poisson noise from the source and background sky. The signal to noise ratio of the brightest pixel in the synthetic images is set to $\sim$50, by adjusting the intensity of the S\'ersic source accordingly. No external shear was simulated in the mock data. 

\subsection{Observed lens galaxies}\label{samples}

We analyse three sets of galaxy-galaxy strong lenses.
These include 42 lenses from the SLACS survey \citep{Bolton2008a} that were fitted without significant residuals by \citet[][hereafter E22]{Etherington2022}'s automated pipeline. 
Most are isolated field galaxies.
They were found by searching for high redshift emission lines in the spectra of low-redshift galaxies obtained through a 3\arcsec\ fibre.
They were then imaged by the HST Advanced Camera for Surveys (ACS) in the F814W band, and processed into stacked images with 0.05\arcsec\ pixels.
We also reprocessed these to measure weak lensing, following the procedure described by \citet{Tam2020}, which supersamples the pixels to 0.03\arcsec.
We exclude lenses J1143-0144 and J1420+6019, for which only one exposure was obtained. 

We analyse 15 lenses from the GALLERY survey \citep{Shu2016b} that were modelled by \citetalias{Etherington2022}. 
These are also field galaxies, found by searching for compact Lyman-$\alpha$-emitting source galaxies in spectra with a 2\arcsec\ fibre.
They were imaged with the HST Wide Field Camera 3 (WFC3) in the F606W band and processed, following \cite{Shu2016c}, into 0.04\arcsec\ pixels.
We do not attempt to measure weak lensing shear in these data.

We analyse 4 galaxy-galaxy strong lenses in the outskirts of galaxy clusters, where we expect a 5--15\% true external shear. Before beginning any analysis, we searched archival HST F814W imaging, and selected lenses with multiple imaging of sources that are extended a similar amount as the field lenses. Positions and redshifts of the selected lenses are given in Table~\ref{Table: cluster lenses}. For MACS1149-GGL18 no source redshift has been recorded, where necessary we assume a source redshift of $z_s=1.5$ and test that the results do not change significantly when we change this assumption over a range of source redshifts from 0.5 to 2. 
Two of our selected lenses had been previously modelled by \citet{Desprez2018}, although constrained using only the positions of multiple images, rather than all the pixels. 
We analysed the HST data similarly to the SLACS lenses, except for the `cosmic snail'. 
For that lens alone, we do not measure weak lensing, but use the independent estimate of shear from \citet{Desprez2018}'s model IV of the galaxy cluster, constrained by cluster-scale strong lensing, and shown by \citet{Desprez2018} to be consistent with ground-based measurements of weak lensing.

{\renewcommand{\arraystretch}{1.1}
\begin{table}
\begin{adjustbox}{max width=\textwidth}
\centering 
\begin{tabular}{lrrrr}
\toprule
                  Lens name &          RA &        Dec &  $z_\mathrm{l}$ &  $z_\mathrm{s}$ \\
\midrule
        MACS1149-GGL18 &  $177.410247$ &  $22.352017$ &   $0.544$ &     - \\
        Abell370-GGL19 &   $39.963013$ &  $-1.534783$ &   $0.375$ &     $2.371$ \\
        MACS1149-GGL20 &  $177.402816$ &  $22.436607$ &   $0.544$ &     $1.806$ \\
 RX J2129-GGL1 (snail) &  $322.428780$ &   $0.108071$ &   $0.235$ &     $1.610$ \\
\bottomrule
\end{tabular}
\end{adjustbox}
\caption{Table of parameters for the 4 galaxy-galaxy lenses in the outskirts of clusters, the lens name refers to the cluster in which the lens resides.}
\label{Table: cluster lenses}
\end{table}}

\if{false}
{\renewcommand{\arraystretch}{1.4}
\begin{table*}
\centering 
\begin{tabular}{c | c c | c c | c c | c | c | c | c} 
\hline\hline

\multirow{2}{*}{\shortstack{Lens\\name}} & \multirow{2}{*}{RA} & \multirow{2}{*}{Dec} & \multicolumn{2}{c|}{SIS neighbour} & \multicolumn{2}{c|}{no neighbour} & \multirow{2}{*}{$\phi^\textrm{bulge}$}& \multirow{2}{*}{$\phi^\textrm{disk}$} & \multirow{2}{*}{$\phi^\textrm{BCG}$} & \multirow{2}{*}{$\phi^\textrm{neighbour}$} \\
\cline{4-7}
 & & & $\phi^\textrm{ext}$ & $\phi^\textrm{PL}$ & $\phi^\textrm{ext}$ & $\phi^\textrm{PL}$ &  & & \\
\hline 
RX\,J2129-GGL1 (snail) & 322.4287798  &  0.1080707 & - & - & 39 & 119 & 123 & 41 & 123 & - \\
Abell370-GGL19 &  39.9630128  & -1.5347831 & 42 & 155 & 25 & 166 & 145 & 109 & -95 & $\sim$90 \\
MACS1149-GGL18 & 177.4102473  & 22.3520171 & -  & - & 146 & 107 & 119 & 157 & 96 & -\\
MACS1149-GGL20 & 177.4028161  & 22.4366067 & 43 & 43 & 32 & 31 & 92 & 75 & 80 & $\sim$30\\
\hline 
\end{tabular}
\caption{}
\end{table*}}

The lens systems we analyse here are sub-samples of the SLACS \citep{Bolton2008a} and GALLERY samples \citep{Shu2016b}. Both samples were selected spectroscopically by analysing residual spectra for higher redshift emission lines; for the GALLERY sample the emission line search is restricted to specifically select for compact Lyman-$\alpha$-emitting source galaxies. The search was carried out within a 3\arcsec\ and 2\arcsec\ fibre for SLACS and GALLERY samples, respectively. The sub-samples of lenses were selected by \citet[][hereafter E22]{Etherington2022} to maintain reasonably uniform data quality to facilitate the automated strong lensing analysis which they performed. In this work, we choose to analyse the 42 SLACS and 15 GALLERY lenses which \citetalias{Etherington2022} demonstrated the automated analysis reliably fits the data without residuals.
 
The SLACS sample were observed on the  Hubble Space Telescope (HST) Advanced Camera for Surveys (ACS) in the F814W band, and the GALLERY sample on the HST Wide Field Camera 3 (WFC3) in the F606W band. For the imaging data on which we perform the strong lensing analysis, the data reduction procedure of the SLACS and GALLERY lenses are described in detail by \cite{Bolton2008a} and \cite{Shu2016c}, respectively. The final spatial resolution is 0.05\arcsec\ per pixel for the SLACS images and 0.04\arcsec\ for the GALLERY images. The point spread function (PSF) was determined for both samples using the \texttt{TinyTim} software \cite{Krist1993}. For the imaging data on which we perform the weak lensing analysis, the data reduction procedure was reduced as described by \citep{Tam2020} and the final spatial resolution is 0.03\arcsec.

The weak lensing procedure will be applied to the SLACS lenses for which, unlike the GALLERY sample, higher wavelength imaging is available. We necessarily remove the lenses J1143-0144 and J1420+6019 from the sample, which due to telescope scheduling issues only have one exposure available. As a result, the final reduced images are not deep enough to detect sufficient background sources for the weak lensing analysis. The remaining 39 SLACS lenses make up our weak lensing sample.

\fi
   
\section{Methods}\label{Method}
\subsection{Weak lensing analysis}
We identified galaxies on lines of sight adjacent to strong lenses using \texttt{SExtractor} \citep{sextractor}, and measured their shapes using the \texttt{PyRRG} \citep{Harvey2019} implementation of the \cite{Rhodes2000} shear measurement method. 
%This measures the apparent ellipticity and shear susceptibility of every galaxy from its weighted multipole moments, correcting them for blurring by the point spread function (PSF) using the \texttt{TinyTim} model \cite{Krist1993}. More details of these procedures are given by \cite{Tam2020}. 
This estimates the mean reduced shear in a patch of sky, by averaging galaxies' apparent shapes 
\begin{equation}\label{eq: WL ellipticity}
    \epsilon_i = \epsilon_i^\mathrm{int} + G\gamma_i,
\end{equation}
which have been transformed by weak lensing (Section~\ref{weak lensing theory}) from an unknown intrinsic shape $\epsilon^\mathrm{int}$. The `shear responsivity' $G$ varies as a function of galaxy flux, and its overall scaling has been calibrated on simulated images with known shear \citep{Leauthaud2007a}. %\citep{Rhodes2000}. 
Under the assumption that galaxies' intrinisic shapes are randomly oriented, i.e.\ $\langle\epsilon_i^\mathrm{int}\rangle\approx0$,
\begin{equation}\label{eq: WL shear estimator}
    \langle \gamma_i\rangle 
    = \left\langle\frac{\epsilon_i - \epsilon_i^\mathrm{int}}{G}\right\rangle
    \approx \left\langle\frac{\epsilon_i}{G}\right\rangle - \left\langle\frac{\epsilon_i^\mathrm{int}}{G}\right\rangle
    \approx \frac{\langle\epsilon_i\rangle}{\langle G \rangle}.
\end{equation}
Following \cite{Massey2007c}, we assume that the median redshift of the lensed galaxies is $z\sim1.26$. Thereafter, following \cite{Smail1994}, we treat them all as being at this effective redshift.
None of our results change significantly if we adjust this value.

We average weak lensing shear measurements from the $\sim$140 galaxies within 60\arcsec\ of the strong lens galaxy (no weights are applied to the galaxies that are averaged). 
The precision of this measurement is limited by the randomness in the distribution of the intrinsic shapes 
\begin{equation}
    \sigma_\mathrm{int}^2=\left\langle\left(\frac{\epsilon_i^\mathrm{int}}{G}\right)^2\right\rangle.
\end{equation}
We measure $\sigma_\mathrm{int}\sim0.3$, consistent with \cite{Leauthaud2007a}, and hence uncertainty $\sigma_\mathrm{int}/\sqrt{140}\approx0.02$ on each component of mean shear. This is similar to uncertainty on our strong lensing measurements. None of our results change significantly if we use a 45\arcsec\ or\ 90\arcsec\ aperture instead. 

Although the line-of-sight directly through each galaxy-scale lens is \textit{not} in the weak lensing regime, we assume that $\langle g_i\rangle\approx\langle\gamma_i\rangle$ still holds since the vast majority of adjacent lines of sights will be only weakly lensed. Nor do we attempt to model and subtract the weak shear due to the galaxy-scale lens itself. Doing so would mix the weak lensing and strong lensing analyses; and it is unnecessary at our achieved level of precision because the near-circular symmetry of most lenses means that the lens contributes negligibly to $\langle\gamma_i\rangle$ inside a 60\arcsec\ circular aperture.

\subsection{Strong lensing analysis}

We analysed all data using the automated strong lens modelling software \texttt{PyAutoLens}\footnote{The \texttt{PyAutoLens} software is open source and available from \url{https://github.com/Jammy2211/PyAutoLens}} \citep{Nightingale2018, Nightingale2021}. This fits parameters of the lens model using all of the pixels in an image (not just e.g.\ locations of the centre of light, as in previous works).

The pipelines used to fit the mock and observed data are described fully in \citetalias{Cao2022} and \citetalias{Etherington2022} respectively.
Briefly, we model the distribution of \textit{mass} in both mock and real data using an elliptical power law (equation~\ref{eq: power law}) plus external shear (equation~\ref{eq: shear strong}). 
We then repeat the fit, fixing external shear $\gamma^{\mathrm{ext}}=0$.
We model the distribution of \textit{light} in in real lens galaxies using a double S\'ersic profile with a centre that is free to vary independently to that of the mass distribution, and for the source galaxy using an adaptive Voronoi mesh of pixels. 
For the mock data, we use \citetalias{Cao2022}'s fit in which the lens light is perfectly subtracted and the source light is modelled as an elliptical S\'ersic. 
\citetalias{Cao2022} also performed fits using a Voronoi mesh for the source light. However, since the mock data were created assuming a S\'ersic source, the model we chose can perfectly describe the source, so  any systematics we observe will be solely due to mismatch between the model and truth of the mass distribution, which is the point of interest in this study.

In \cref{AppA} we perform two test of our shear measurements: (i) we refit every strong lens model including one or two nearby bright galaxies explicitly as singular isothermal sphere and; (ii) use a different source analysis which assumes a Delaunay mesh. We recover consistent shear magnitudes and position angles across the lens sample. Results presented in the main paper therefore cannot be explained as due to lens models missing line-of-sight galaxies or systematics in the source analysis.

%We repeat these analyses, but without free parameters for external shear, i.e.\ light is deflected only by the power-law distribution of mass in the lens galaxy. The procedure for this analysis applied to the observed and image data is identical to the pipelines described above but with $\gamma^{\mathrm{ext}}=0$.

\subsection{Multipole perturbations of an ellipse}
\label{sec:multipole_method}

\begin{figure}
    \centering
    \includegraphics[width=0.8\linewidth]{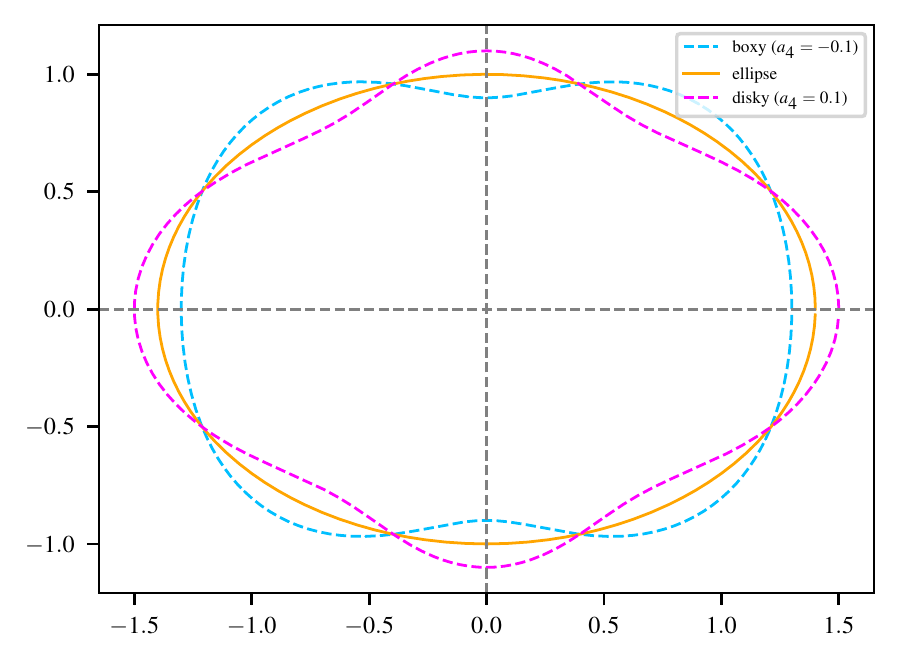}
    \caption{Examples of boxy ($a_4=-0.1$, blue dashed curve) and disky ($a_4=+0.1$, pink dashed curve) perturbations to an ellipse (orange curve). The perturbations shown are $\sim$10 times larger than those typically observed. In both cases the perturbation at $45^\circ$, $b_4=0$ (see equation~\ref{eq: delta r}). %\rjm{Make sure you use two colours that still look different from each other when printed in back \& white. Make one dashed and the other dotted?}
    }
    \label{Figure: boxy and disky}
\end{figure}

In Section~\ref{sec:boxiness_disciness} we shall investigate whether the strong lensing external shears depend on deviations from an elliptically symmetric distribution of mass. Specifically, we shall quantify the \textit{multipole} deviations of two types of contour: the iso-convergence contour at $\kappa=1$ of the gNFW+MGE distributions used to create mock data, and the critical curves of both the mock and the observed galaxies. These contours are stored as a 2D array of points in polar coordinates $[\phi_\mathrm{contour}, R_\mathrm{contour}]$. We calculate perpendicular deviations of each point from the true ellipse
\begin{equation}
    R_{\mathrm{el}}(\phi) = \frac{a\sqrt{1-\epsilon^2}}{\sqrt{1-\epsilon^2\cos^2(\phi-\phi_{\mathrm{el}})}},
\end{equation}
where $a$ is the major axis, $\phi_{\mathrm{el}}$ is the major axis orientation, and $\epsilon$ is the eccentricity (defined as $\epsilon^2\equiv1-b^2/a^2$ where $b$ is the minor axis). The deviations are then parameterised using multipoles 
\begin{equation}\label{eq: delta r}
    \delta R_m (\phi;a_m, b_m) = \sum a_m \cos(m(\phi-\phi_{\mathrm{el}})) + b_m \sin(m(\phi-\phi_{\mathrm{el}})) \, ,
\end{equation}
where $m$ is the order of the multipole, and $a_m$ and $b_m$ are the magnitude of the deviations with symmetry along or at $45^\circ$ to the major and minor axes, respectively. We then perform a non-linear search to fit the model 
\begin{equation}\label{multipole model}
    R(\phi;a_4,b_4) = R_{\mathrm{el}}(\phi) + \delta R_4 (\phi;a_4,b_4)
\end{equation}
to the radial values of the contour. We assume uniform priors on the free parameters in the fit over a reasonable range and fit for them using the nested sampling algorithm \texttt{dynesty}. We assume the residual errors can be described by a Gaussian distribution and maximise the likelihood
%\begin{equation}
%    \mathscr{L}(R, \sigma |R_i) =  \prod_{i}\frac{\mathrm{exp}\Big[-\frac{(R(\phi_i)-R_i)^2}{2\sigma^2}\Big]}{\sqrt{2\pi\sigma^2}},
%\end{equation}
\begin{equation}
    \mathscr{L}(R |R_i, \sigma) =  \prod_{i} \left[ \frac{1}{\sqrt{2\pi\sigma^2}}~
    \mathrm{exp}\left(-\frac{\left(R(\phi_i)-R_i\right)^2}{2\sigma^2}\right)\right],
\end{equation}
where $R_i$ are the radial values of the contour and $R(\phi_i;a_4,b_4)$ are the model predicted values from equation~\eqref{multipole model} at each angular coordinate in the contour $\phi_i$. Curves with best-fit values of $a_4>0$ are `disky'; those with $a_4<0$ are `boxy' (see Figure~\ref{Figure: boxy and disky}).

%The uncertainty $\sigma$ on the contour's radial values is ill-defined since, at least in the case of the mock sample, the values are just known quantities. We therefore set this \rjm{explain what is actually now done} arbitrarily to 0.02 which is approximately equal to the pixel scale of the sub-grid on which the contours are calculated.
   
\begin{figure}
    \centering
    \includegraphics[width=0.8\linewidth]{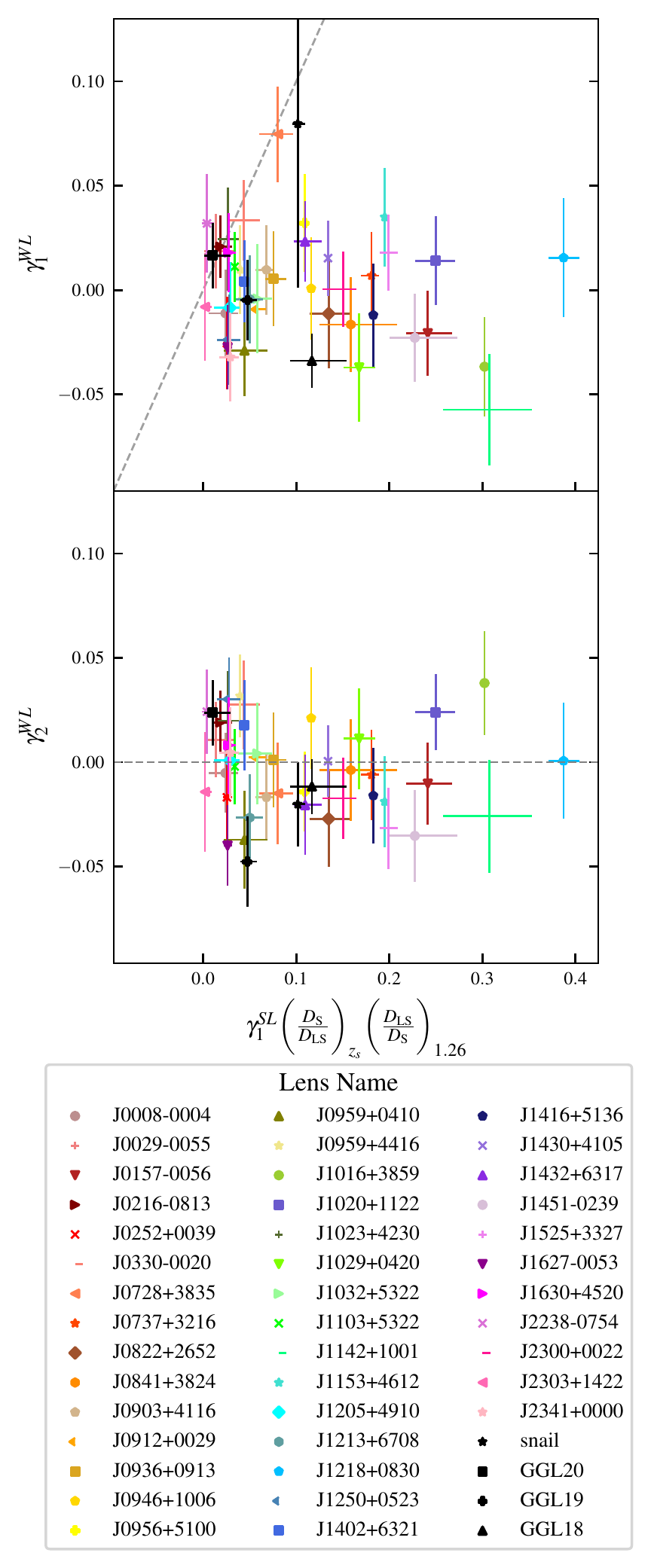}
    \caption{Values of the shear along the lines of sight to 39 galaxy-galaxy lenses, independently measured using strong lensing `external' shear $\gamma^{\mathrm{SL}}$ and weak lensing $\gamma^{\mathrm{WL}}$. Shears are oriented such that $\gamma^{\mathrm{SL}}_2=0$, and rescaled to be at the same effective redshift. If strong and weak lensing shears were identical, all points would lie on the dashed lines. We instead find that external shears inferred from strong lensing are consistently larger than those measured by weak lensing, and not aligned.}
    \label{Figure: wl shear comp}
\end{figure}

\section{Results}\label{Results}

\subsection{SL shears do not correlate with WL shears}\label{wl comp}

Strong lensing measurements of shear $\gamma^\mathrm{SL}$ (obtained as the best-fit external shear $\gamma^\mathrm{ext}$) typically have amplitudes up to an order of magnitude larger than weak lensing measurements $\gamma^\mathrm{WL}$ along the same line of sight. 
The mock lenses have mean best-fit $|\langle\gamma^\mathrm{SL}\rangle|=0.019\pm0.002$, despite the true values all being $\gamma^\mathrm{ext}=0$.
Our measurement using a PL+ext mass model is consistent with \citetalias{Cao2022}'s value $|\langle\gamma^\mathrm{SL}\rangle|=0.015$ using pixel-based source reconstructions. 
The real lenses have mean best-fit shear $|\langle\gamma^\mathrm{SL}\rangle|=0.098\pm0.011$, which is much larger than both the measured weak lensing shear $|\langle\gamma^\mathrm{WL}\rangle|=0.028\pm0.002$.

Strong lensing measurements of shear do not correlate with weak lensing measurements (Figure~\ref{Figure: wl shear comp}). 
To make this comparison (for the real lenses only), we first define rotated coordinate systems such that $\gamma_1^\mathrm{SL}=\gamma^\mathrm{SL}\geq 0$ and $\gamma_2^\mathrm{SL}=0$. Thus we need plot only three of the four components of shear.
Second, we compensate for the different redshifts of the strongly lensed and weakly lensed sources by rescaling values of $\gamma_1^\mathrm{SL}$ by $(D_{\rm s}/D_{\rm ls})_{z_s'=z_s}(D_{\rm ls}/D_{\rm s})_{z_s'=1.26}$, i.e.\ the effective value at the redshift of the weakly lensed galaxies (see eqn~\ref{eq:phi(Sigma)}). 
This scaling is exact only if the external shear is both real and dominated by neighbouring structures at the same redshift as the lens (\citealt{Wong2011} found that 5/8 of the shear is from neighbouring structures).
In any case, the rescaling is by a factor with mean of only 1.26 and rms 1.06, 
% (minimum 0.86 and maximum 2.00).
and our conclusions do not change if the rescaling is omitted or normalised to a different redshift. 
%Note that strongly lensed and weakly lensed sources may be at different redshifts. However, for a fixed lens (mass distribution and redshift), $\gamma_i\propto D_{\rm ls}/D_{\rm s}$, so %the quantity $\gamma_i D_{\rm s}/D_{\rm ls}$
%\begin{equation}
%\gamma_i^\infty \equiv \frac{D_{\rm s}}{D_{\rm ls}}\,\gamma_i 
%\end{equation}
%(i.e.\ the shear experienced by a source at infinite distance) should be constant, regardless of how it is measured.
If strong and weak lensing measure the same quantity, we then expect $\gamma_1^\mathrm{WL}$ to correlate with $\gamma_1^\mathrm{SL}$, and $\gamma_2^\mathrm{WL}$ to scatter around zero.
We find that $\langle\gamma_2^\mathrm{WL}\rangle =-0.004\pm0.003$ is on average below zero, and its scatter (0.02) is consistent with uncertainties calculated from the distribution of weak lensing shears.
The best-fit slope $\gamma_1^\mathrm{WL}=(-0.06\pm0.04)\gamma_1^\mathrm{SL}$ actually infers a negative correlation, however this does not take into account the uncertainty on the strong lensing shears and so the uncertainty is likely underestimated. The Pearson correlation coefficient $-0.19\pm0.22$ implies that, if there is a correlation, there is also a large amount of scatter.

There are eight lenses for which $\gamma^\mathrm{ext}\approx\gamma^\mathrm{WL}$, including two of the four lenses which reside in the outskirts of clusters (see Section~\ref{Sec: cluster lesnes} for further discussion). However, there does not appear to be anything unique about these lenses that would make the shear possible to measure in these cases.
%
% FOR 0912 the mass axis ratio is 0.79 with position angle 27degrees, and it's 81degrees offset from the shear, so roughly anti-aligned
%the two sersics are similar to the mass distribution, axis ratios of 0.74, and 0.61, and position angles 14 and 11 degrees

\subsection{SL shears are (suspiciously) aligned with the mass}\label{external shear behaviour}

\begin{figure}
\centering
    \begin{subfigure}{\columnwidth}
    \centering
    \includegraphics{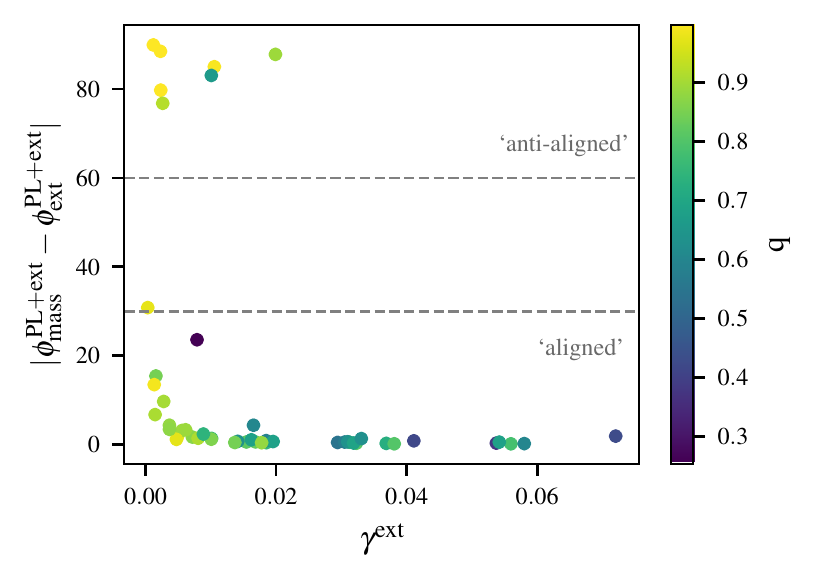}
    \end{subfigure}\label{Figure: shear angle q mocks}
    \begin{subfigure}{\columnwidth}
    \centering
    \includegraphics{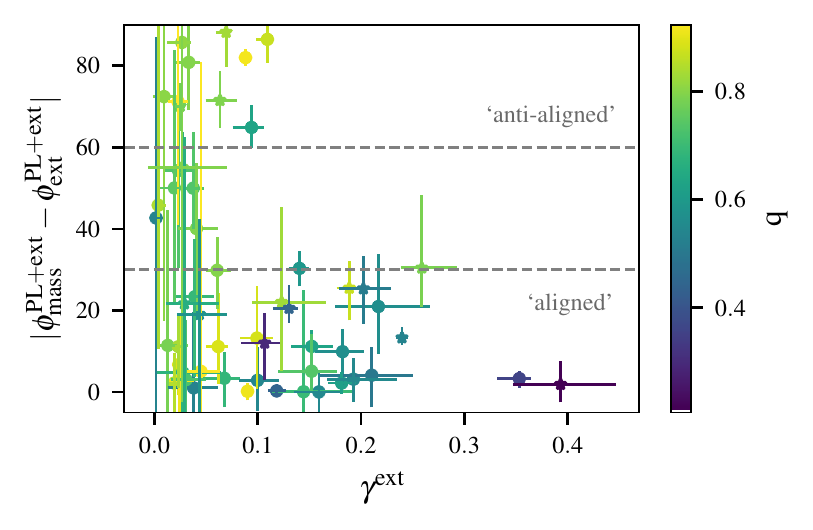}
    \end{subfigure}\label{Figure: shear angle q data}
    \caption{Relative orientation of the strong lensing `external' shear and the major axis of the lens mass, in mock ({\it top panel}) and real HST ({\it bottom panel}) data. 
    In both cases, most of the shears are suspiciously aligned ($\phi^\mathrm{PL+ext}_\mathrm{mass}-\phi^\mathrm{PL+ext}_\mathrm{ext}\leqslant30^\circ$) or anti-aligned ($\phi^\mathrm{PL+ext}_\mathrm{mass}-\phi^\mathrm{PL+ext}_\mathrm{ext}\geqslant60^\circ$) with the lens mass distribution. If the $\gamma_\mathrm{ext}$ parameter were measuring true external perturbations, the orientations would be random. 
    Points are coloured by the best-fit axis ratio of the lens mass distribution. Highly elliptical lenses often lead to high values of $\gamma_\mathrm{ext}$.}
    \label{Figure: shear angle q}
\end{figure}

For both mock and real data, the best-fit `external' shear is usually aligned with the major axis of the lens mass ($\phi^\mathrm{PL+ext}_\mathrm{mass}-\phi^\mathrm{PL+ext}_\mathrm{ext} \leqslant 30^\circ$) or with its minor axis ($\phi^\mathrm{PL+ext}_\mathrm{mass}-\phi^\mathrm{PL+ext}_\mathrm{ext}\geqslant60^\circ$): see figure~\ref{Figure: shear angle q}. If external shears were truly measuring external perturbations, their orientations would be random \citep[modulo intrinsic alignments between the shape of a galaxy and its surrounding tidal field, but these are much smaller than our achieved measurement precision;][]{Zhang2022}. The preference for aligning with the mass distribution again suggests an `internal' shear that compensates for the inability of a power law model to represent the more complex true distribution of mass.
Furthermore, the highest values of $\gamma_\mathrm{ext}$ are also usually found in the most elliptical lenses.

In mock data, 84$\%$ of external shears are aligned with the mass distribution: their mean offset is $3^\circ$ with an rms scatter of $5^\circ$. 14$\%$ of external shears are anti-aligned with the mass distribution, with a mean of $85^\circ$ and scatter of $6^\circ$. Only one lens has a best-fit external shear that is neither aligned nor anti-aligned, but this also has the lowest shear amplitude ($\gamma_\mathrm{ext}=0.0003$), so the angle $\phi_\mathrm{ext}$ is ill-defined. 
The Pearson correlation coefficient between the best-fit axis ratios and external shears is $-0.63$. 

{\renewcommand{\arraystretch}{1.5}
\begin{table*}
\centering
\begin{adjustbox}{max width=\textwidth}

\begin{tabular}{l | c c c c | c c c c | c c c c | c | c } 
\hline\hline

\multirow{2}{*}{Lens name} & \multicolumn{4}{c|}{SIS neighbour} & \multicolumn{4}{c|}{no neighbour}  & \multirow{2}{*}{$\gamma_1^\textrm{WL}$}& \multirow{2}{*}{$\gamma_2^\textrm{WL}$} & \multirow{2}{*}{$\gamma^\textrm{WL}$}& \multirow{2}{*}{$\phi^\textrm{WL}$} & \multirow{2}{*}{$\phi^\textrm{BCG}$} & \multirow{2}{*}{$\phi^\textrm{neighbour}$} \\
\cline{2-9}
 & $\gamma^\textrm{ext}$ & $\phi^\textrm{ext}$ & $q^\textrm{PL}$ & $\phi^\textrm{PL}$ & $\gamma^\textrm{ext}$ & $\phi^\textrm{ext}$ & $q^\textrm{PL}$ & $\phi^\textrm{PL}$ & & & & & \\
\midrule
        MACS1149-GGL18 &  $0.13^\textrm{+0.04}_\textrm{-0.02}$ &  $40^\textrm{+5}_\textrm{-5}$ &  $0.48^\textrm{+0.04}_\textrm{-0.02}$ &  $147^\textrm{+9}_\textrm{-8}$ &  $0.24^\textrm{+0.02}_\textrm{-0.02}$ &    $26^\textrm{+2}_\textrm{-2}$ &  $0.82^\textrm{+0.02}_\textrm{-0.02}$ &  $174^\textrm{+19}_\textrm{-22}$ &  $0.01^\textrm{+0.01}_\textrm{-0.01}$ &    $-0.04^\textrm{+0.01}_\textrm{-0.01}$ &          0.04 &   -40 &         85 &           95 \\
        Abell370-GGL19 &  $0.06^\textrm{+0.01}_\textrm{-0.01}$ &  $40^\textrm{+3}_\textrm{-6}$ &  $0.92^\textrm{+0.01}_\textrm{-0.01}$ &   $23^\textrm{+1}_\textrm{-1}$ &  $0.07^\textrm{+0.03}_\textrm{-0.02}$ &   $34^\textrm{+7}_\textrm{-14}$ &  $0.72^\textrm{+0.03}_\textrm{-0.02}$ &     $26^\textrm{+5}_\textrm{-6}$ &  $0.05^\textrm{+0.02}_\textrm{-0.02}$ &    $-0.01^\textrm{+0.02}_\textrm{-0.02}$ &          0.05 &    -8 &       -100 &           35 \\
        MACS1149-GGL20 &                                     - &                             - &                                     - &                              - &  $0.01^\textrm{+0.02}_\textrm{-0.01}$ &  $13^\textrm{+47}_\textrm{-49}$ &  $0.51^\textrm{+0.02}_\textrm{-0.01}$ &    $106^\textrm{+1}_\textrm{-2}$ &  $0.01^\textrm{+0.02}_\textrm{-0.02}$ &     $0.04^\textrm{+0.02}_\textrm{-0.02}$ &          0.04 &    40 &        -83 &            - \\
 RX J2129-GGL1 (snail) &                                     - &                             - &                                     - &                              - &  $0.11^\textrm{+0.01}_\textrm{-0.01}$ &   $35^\textrm{+2}_\textrm{-2}$ &  $0.93^\textrm{+0.01}_\textrm{-0.01}$ &   $-61^\textrm{+10}_\textrm{-12}$ &  $0.08^\textrm{+0.08}_\textrm{-0.08}$ &  $-0.02^\textrm{+-0.02}_\textrm{--0.02}$ &          0.08 &    28 &        -57 &            - \\
\bottomrule
\end{tabular}
\end{adjustbox}
\caption{Summary of strong and weak lensing parameters for the 4 galaxy-galaxy lenses that reside in clusters. All angles are in degrees anticlockwise from West.}
\label{Table: cluster orientations}
\end{table*}}

Real HST data produce a similar pattern, but with more scatter. 
Best-fit `external' shears are aligned with the mass distribution in $68\%$ of the lenses, and anti-aligned in $20\%$.
All the remaining $12\%$ have $\gamma_\mathrm{ext}<0.04$, so the angles $\phi^\mathrm{PL+ext}_\mathrm{ext}$ are noisy.
The Pearson correlation coefficient between the best-fit axis ratios and external shears is $-0.60$. 
Despite the inferred external shears being an order of magnitude larger in the observations than in the mock data, the mass distributions are similarly elliptical: $\langle q\rangle=0.77\pm0.02$ in the mocks, and $0.69\pm0.02$ for HST data.

\subsection{SL Shear Amplitudes Are Too Large}

Along random lines of sight through the universe a shear amplitude of $\sim1-3\%$ is expected \citep{Keeton1997}, even accounting for the different scales on which they are averaged \citep{Valageas2004, Wong2011}. SLACS and GALLERY strong lenses typically exist in field environments where shear amplitudes should therefore never exceed $\sim 5\%$ \citep{Treu2009}, however figure~{Figure: shear angle q} shows the majority of measured SL shear amplitudes exceed $\sim5\%$ and a large fraction exceed $\sim10\%$. Therefore, even without WL data or detailed inspection of the SL mass models, the SL shear measurements alone are indicative of a non external origin.

\subsection{Lenses in clusters}\label{Sec: cluster lesnes}
\subsubsection{RX J2129-GGL1 (snail)}
Of the galaxy-galaxy lenses in clusters the snail measures shears that agree most closely between the two methods. Notably, it measures the largest shear independent of the strong lensing method. Although this measurement was constrained by cluster scale strong lensing, \cite{Desprez2018} demonstrated that this value of shear is in agreement with that derived from weak lensing analysis of CLASH data (Figure 13 of that paper). The strong lensing external shear is anti-aligned with it's mass distribution, although this is expected since the mass distribution is coincidentally orientated with it's major axis pointing towards the cluster centre (Table~\ref{Table: cluster orientations}). 

\begin{figure*}
    \begin{subfigure}{\columnwidth}
    \includegraphics[width=0.9\columnwidth]{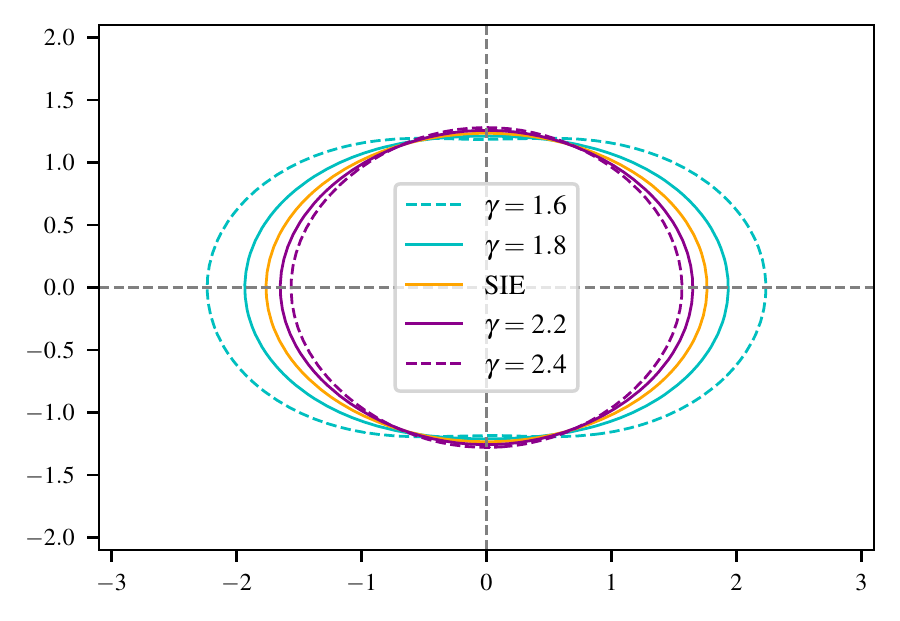}
    \end{subfigure}
    \begin{subfigure}{\columnwidth}
    \includegraphics{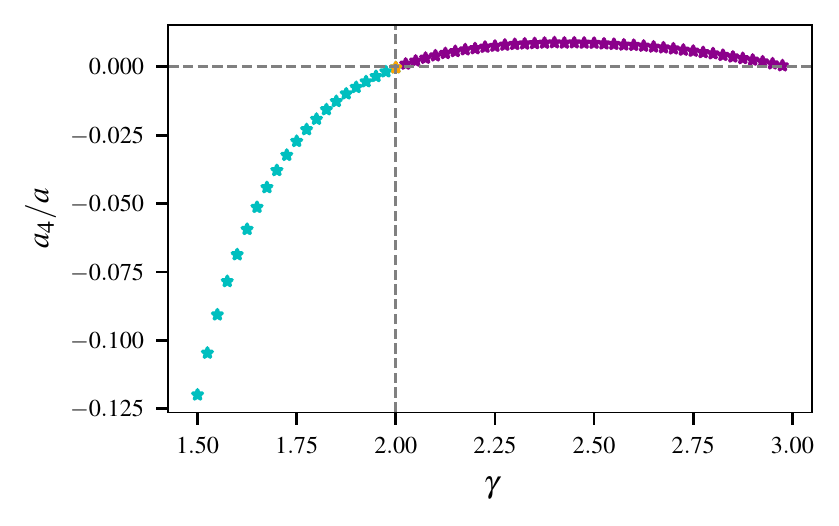}
    \end{subfigure}
    \begin{subfigure}{\columnwidth}
    \includegraphics[width=0.9\columnwidth]{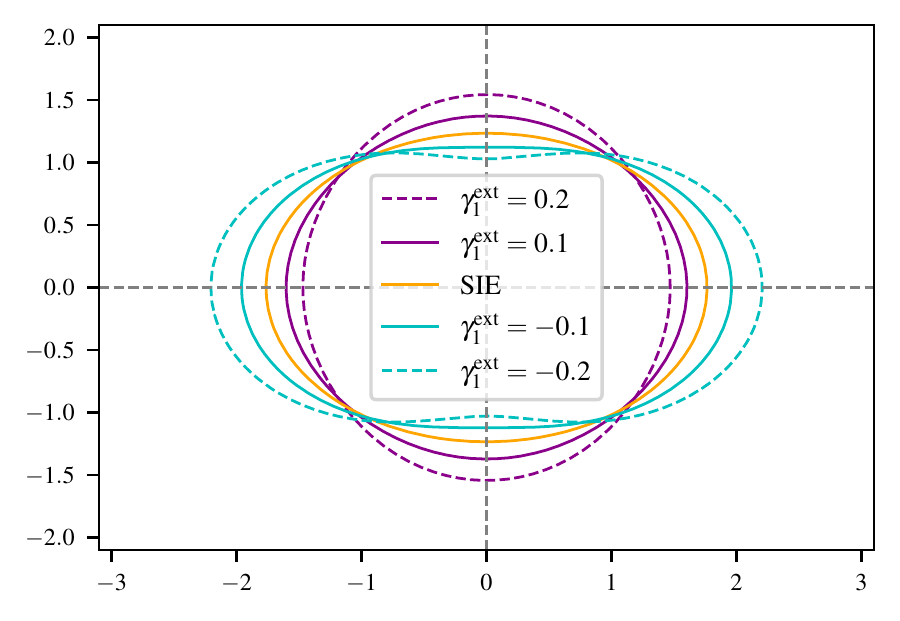}
    \end{subfigure}
    \begin{subfigure}{\columnwidth}
    \includegraphics{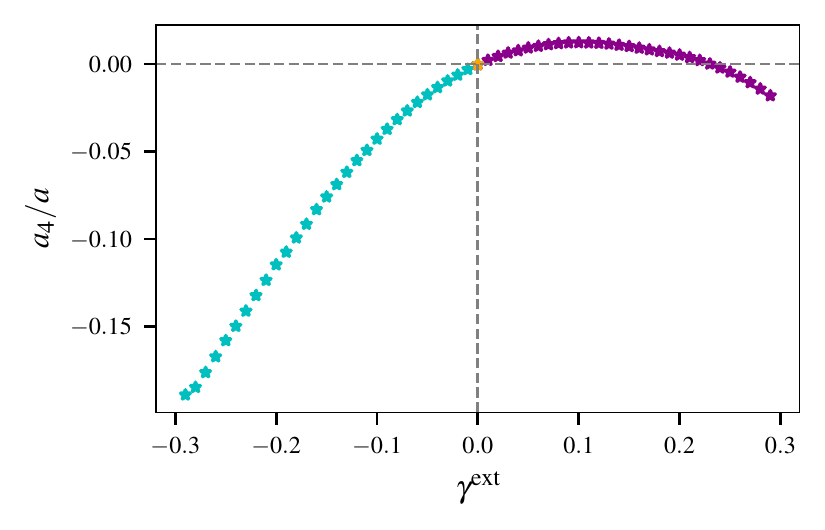}
    \end{subfigure}
    \caption{What causes boxiness or diskiness? A Singular Isothermal Elliptical (SIE) mass distribution with a horizontal major axis has critical curves that are also elliptical with a horizontal major axis (orange). 
    The critical curves are perturbed if the slope of the density profile $\gamma\ne 2$ (top left panel) or the external shear $\gamma^\mathrm{ext}\ne 0$ (bottom left panel). 
    In particular, an aligned shear ($\gamma_1^\mathrm{ext}>0$) stretches the critical curves vertically (and the image horizontally); an anti-aligned shear ($\gamma_1^\mathrm{ext}<0$) does the opposite. 
    Multipole measurements $a_4/a$ of the critical curve are shown as a function of slope (top right panel) and external shear (bottom right panel), where $a_4/a>0$ is ``disky'' and $a_4/a<0$ is ``boxy''.}
    \label{Figure: critical curve perturbations}
\end{figure*}

\subsubsection{MACS1149-GGL20}
The shears measured using the independent probes for MACS1149-GGL20 are also in agreement, although the shear magnitude is much lower than is measured for the snail. In fact, this is one of the few lenses that measures a lower best-fit value of shear magnitude with strong lensing $\gamma^\mathrm{ext}=0.01$ than it does with the weak lensing method $\gamma^\mathrm{WL}=0.04$. \citet{Desprez2018} found that measurements of external shear from modelling the GGL alone were underestimated compared to the shears constrained using a full scale model of the galaxy cluster. However, both of these measurements are larger than either of the shears measured in this work. The authors measure an external shear magnitude $\gamma^\mathrm{ext}=0.13^{+0.08}_{-0.06}$ when modelling the potential of the lens as a double Pseudo-Isothermal Elliptical profile (dPIE), significantly larger than that measured in this work $\gamma^\mathrm{ext}=0.01^{+0.02}_{-0.01}$. However, we measure a more elliptical mass distribution $q=0.51^{+0.02}_{-0.01}$ than was constrained by \citep{Desprez2018} $q=0.81$. The degeneracy between shear and axis ratio may therefore explain this discrepancy. As with the snail, the mass distribution coincidentally points towards the cluster centre, therefore the anti-alignment of the external shear with the lens' mass distribution that we infer is to be expected. 

\subsubsection{Abell370-GGL19}
We measure a similar shear magnitude with strong lensing for Abell370-GGL19 as we do with weak lensing, but the strong lensing external shear is suspiciously orientated towards a nearby neighbour galaxy and is aligned with the mass distribution (see Table~\ref{Table: cluster orientations}). We therefore repeat the fit including free parameters for a singular isothermal sphere (SIS; $\gamma=2$ and $q=1$ in equation~\ref{eq: power law}) fixed at the centre of the neighbour galaxy. The results including the mass of the neighbour galaxy do not change significantly (see the SIS neighbour column of Table~\ref{Table: cluster orientations} compared to the no neighbour column), although the power-law mass distribution does become less elliptical. 

\subsubsection{MACS1149-GGL18}
There is also a neighbour galaxy in close proximity to MACS1149-GGL18. We, therefore repeat the fit for including an SIS as was done for Abell370-GGL19. The shear is significantly overestimated as compared with weak lensing when the neighbour is not included in the fit. Including the neighbour galaxy halves the inferred strong lensing external shear, but this is still inconsistent with the weak lensing measurement. The shear is anti-aligned with the power-law mass distribution which, given that the mass distribution is not aligned with the cluster galaxy in this case, suggests the external shear may be acting internally as discussed in the previous section.

\section{Analysis}\label{section: analysis}

\subsection{External shear may compensate for boxiness/diskiness}
\label{sec:boxiness_disciness}

Our measurements in Section~\ref{Results} suggest that $\gamma^\mathrm{ext}$ mostly just compensates for the inability of a power law model to capture the complex distributions of mass (gNFW+MGE for the mocks, and likely more complex for real galaxies).
This is consistent with the conclusion of \cite{Keeton1997}, who inferred that the $\langle\gamma^\mathrm{ext}\rangle\sim10$--$15\%$ required to fit point-source quad lenses, must reflect an inability of the lens model to capture a complex distribution of mass: perhaps misalignment between light and dark matter. \cite{Witt1997} reached a similar conclusion, and derived an analytical prediction of the shear required by an elliptical potential to fit quad lenses. 

If the external shears result from a lack of complexity in the power law model to describe the underlying distribution of mass, one can ask what type of complexity the data requires. One possible deviation from the symmetry of an elliptical power law is boxiness and diskiness (see Section~\ref{sec:multipole_method}). We shall now investigate whether spurious external shear could arise to compensate for boxy/disky lens galaxies.

\subsubsection{External shear creates boxy/disky critical curves}
\label{sec:shear_critical_curves}
An isothermal elliptical mass distribution has an elliptical critical curve (oriented in the same direction as the mass distribution but at $90^\circ$ to the elongation of light from the source galaxy; \citealt{saasfee}). However, changing the power law slope, $\gamma\ne2$, or adding an external shear, $\gamma^\mathrm{ext}\ne 0$, perturbs the critical curves (left panel of Figure~\ref{Figure: critical curve perturbations}). These perturbations include significant $a_4/a$ moments (right panels of Figure~\ref{Figure: critical curve perturbations}), although they visually appear to be more than pure $m=4$ modes (c.f.\ Figure~\ref{Figure: boxy and disky}).

\begin{figure}
    \centering
    \includegraphics[width=\linewidth]{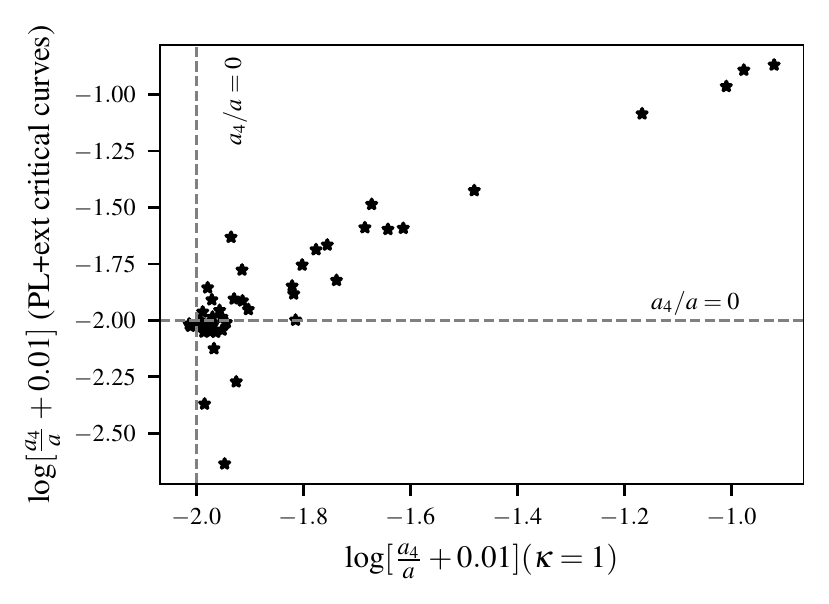}
    \caption{In mock lenses, disky ($a_4/a>0$) perturbations of the $\kappa=1$ isodensity contour correlate with disky perturbations of the critical curves -- which can also be measured for real lenses. To better visualise the correlation, values have been transformed by $\log[a_4/a+0.01]$, with dashed lines indicating $a_4=0$. 
    Unfortunately, the mock data do not include any lenses with significantly boxy ($a_4/a<0$) distributions of mass.
    These points come from fits with Sersic sources, so the uncertainties are not comparable to those from analyses with pixellated sources.
    }
    \label{Figure: a4 comp}
\end{figure}

\subsubsection{Disky critical curves come from disky mass distributions}
\label{sec:crit_curve_mass_distribution}

The distributions of mass in our mock lenses happen to be almost all disky. We could measure any isodensity contour, but the $\kappa$=$1$ contour will be near the most sensitive region for lens fitting. These iso-convergence contours have mean $\langle|a_4/a|\rangle=0.01$ and $\langle|b_4/a|\rangle=0.0005$. Only three lenses are boxy, but not usefully so, with very low values $\langle a_4/a\rangle=-0.0003$.

Critical curves of the best-fit PL+ext models to our mock data show $a_4/a$ moments highly correlated with those of the density contours (Figure~\ref{Figure: a4 comp}). Again, $\langle|b_4/a|\rangle=0.0001$ is an order of magnitude lower. Studying the same mocks, \citetalias{Cao2022} also noted that `external' shear allowed the best-fit critical curves to better match the true critical curves. % (and since the mocks were created without $\gamma^\mathrm{ext}$, they dubbed this `internal' shear).
We find two systems with boxy critical curves $a_4/a < -0.01$. Subject to some scatter, however, we conclude that the diskiness of isodensity contours and critical curves are highly correlated.

\begin{figure}
    \centering
    \includegraphics[width=\linewidth]{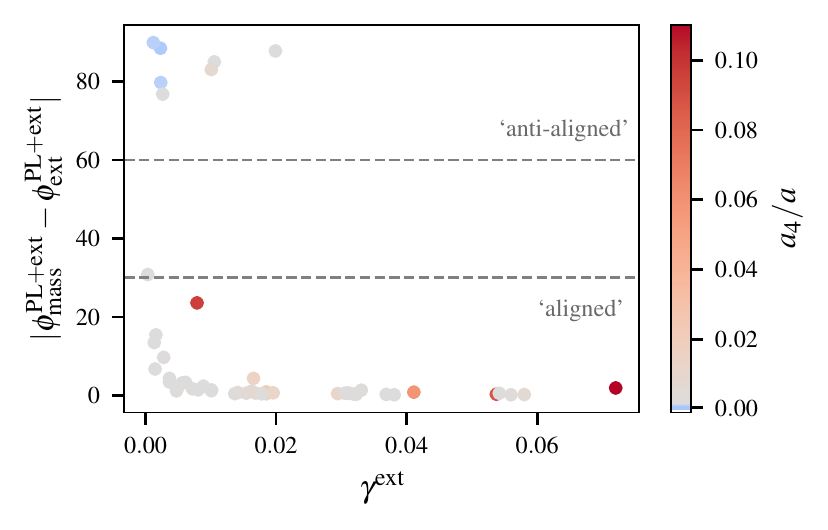}
    \caption{For mock lenses that were simulated without external shear. Angle between the best-fit values of external shear and the major axis of the lens mass distribution, $|\phi^\mathrm{PL+ext}_\mathrm{mass}-\phi^\mathrm{PL+ext}_\mathrm{ext}|$ in degrees, as a function of the amplitude of the best-fit external shear, $\gamma_\mathrm{ext}$. Points are coloured by the magnitude of the inferred critical curves deviation from elliptical symmetry $a_4/a$, values of $a_4/a<0$ correspond to boxy critical curves and $a_4/a>0$ to disky ones. Note the red and blue colour points are orders of magnitude different, the blue range reaches a maximum of 0.001.
    }
    \label{Figure: shear angle boxiness mocks}
\end{figure}

Notably, all mock lenses whose best-fit external shear is aligned with the mass distribution have very disky critical curves (red points in Figure~\ref{Figure: shear angle boxiness mocks}), and the three mock lenses with boxy critical curves have anti-aligned shear. Furthermore, $a_4$ typically increases with the external shear (Pearson correlation coefficient 0.45) and with the axis ratio of the lens mass (Pearson correlation coefficient $-0.73$). This may be tentative evidence that (some of) the dichotomy of aligned and anti-aligned shears may be caused by diskiness or boxiness.

\begin{figure}
    \centering
    \includegraphics[width=\linewidth]{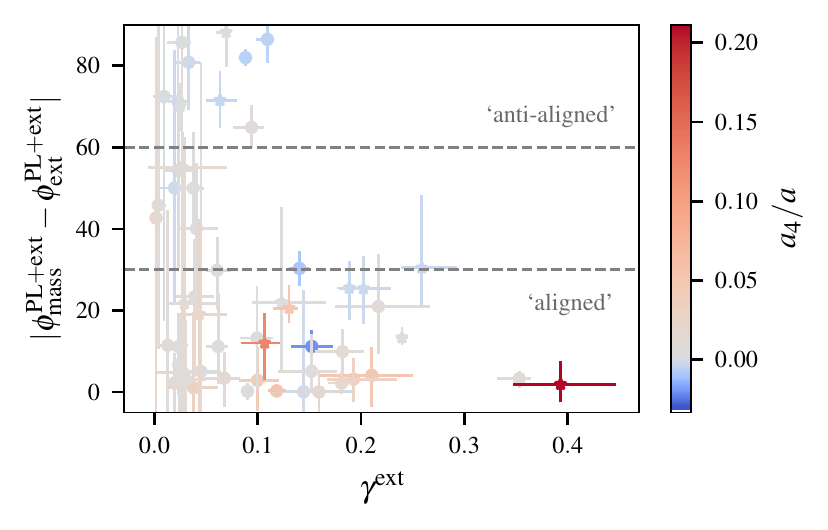}
    \caption{Same as for Figure~\ref{Figure: shear angle boxiness mocks} but for the observed SLACS and GALLERY lenses. The inferred external shears have a similar distribution of aligned and anti-aligned shears as the mock data sample, indicating they too may be acting internally. Note the increase in the scale of shear magnitude $\gamma_\mathrm{ext}$ and elliptical deviations $a_4/a$ compared to the mock data sample.}
    \label{Figure: shear angle boxiness data}
\end{figure}

\subsubsection{Does boxiness/diskiness cause `external' shear?}
\label{sec:diskiness_is_ext_shear}

Scatter in real data is larger than in the mocks. However, for SLACS and GALLERY lenses, the best-fit critical curves have mean $\langle|a_4/a|\rangle=0.016$, similar to the mocks. Most ($79\%$ of) lenses with best-fit `external' shear that is aligned with the mass distribution have disky critical curves; and most ($70\%$) with anti-aligned shear have boxy critical curves (Figure~\ref{Figure: shear angle boxiness data}). Moreover, lenses with the largest amplitude of external shear also have critical curves with the largest deviations from elliptical (Pearson correlation coefficient of 0.48 with $a_4$ and 0.65 with $b_4$).

This provides tentative evidence that `external' shear in typical lensing analyses is really caused by the inability of parametric mass models to capture the complex distribution of mass in a lens. A substantial portion of that complexity may be diskiness or boxiness of the mass distribution. This creates diskiness or boxiness in the critical curves (Section~\ref{sec:crit_curve_mass_distribution}), which leads to a spurious external shear (Section~\ref{sec:shear_critical_curves}).
We have not been able to quantify the relative contributions to external shear from true shear, diskiness/boxiness, or other sources. %Indeed, the full complexity of real lenses means that there are probably several other sources.

\subsubsection{More things probably cause `external' shear too}
\label{section:twisting}

There are likely more sources of complexity in real mass distributions, which cause (or are compensated by) external shear. These confounding factors would explain the looser correlations and larger scatter than in the mocks. Just the observation that real lenses have external shears with amplitudes six times greater than mocks implies that their distribution of mass deviates more from a power law.

We speculate that the isodensity contours of a lens might be twisted (misaligned as a function of radius), like their isophotoes. Indeed, the critical curves of SLACS and GALLERY lenses have a handedness, with $\langle|b_4/a|\rangle=0.012$ two orders of magnitude larger than the mocks. If the critical curves do tell us something about the distribution of mass, this may indicate twisted isodensity contours. Twisting is also suggested by the inconsistently measured position angle when real data are fitted with and without external shear (see Figure~\ref{Figure:shear comp data}); this is not present in the mocks. \cite{VanDeVyvere2022a} also found that twists in the underlying mass distribution are typically absorbed by changes in orientation of the mass distribution and shear in a PL+ext model. The maximum amplitude of external shear that can be attributed to these twists is around $5\%$ to $8\%$ \citep[see Figures B3 and B4 of][]{VanDeVyvere2022a}, but lower values are more common for most of their mock lenses. Given many of our strong lenses have shear amplitudes above $\sim 10\%$, twists alone cannot explain all of the SL shears measured.

\subsection{External shear biases other SL model parameters}\label{section: shear tests}

Since best-fit values of `external' shear may not (entirely) represent the physical quantity they are imagined to, we now test which other parameters in the mass model are biased by their inclusion, and which are still robustly measured.
As an extreme alternative, we repeat all measurements but fix $\gamma^\mathrm{ext}_1=\gamma^\mathrm{ext}_2=0$ when we refit the mass distribution.

\subsubsection{Mock lenses}
\label{section: shear tests mock}

\begin{figure}
    \centering
    \includegraphics[width=\linewidth]{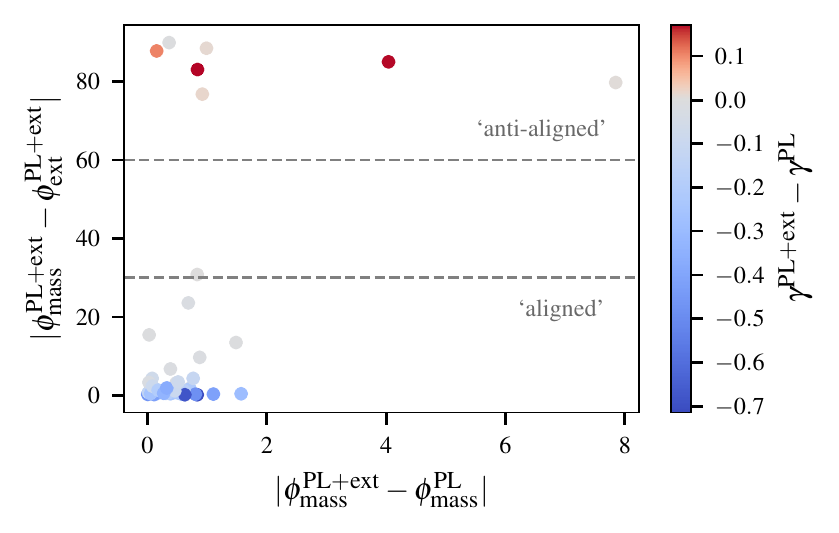}
    \caption{For mock lenses that were simulated without external shear. Angle between the best-fit external shear and the major axis of the lens mass distribution, $|\phi^\mathrm{PL+ext}_\mathrm{mass}-\phi^\mathrm{PL+ext}_\mathrm{ext}|$ in degrees, as a function of the change in orientation of the major axis when fitting with and without an external shear, $|\phi^\mathrm{PL+ext}_\mathrm{mass}-\phi^\mathrm{PL}_\mathrm{mass}|$. Points are coloured by the difference in power-law slope inferred between the models fitted with and without an external shear ($\gamma^\mathrm{PL+ext}-\gamma^\mathrm{PL}$). Systems with best-fit shear aligned to the mass systematically decrease in power-law slope when the external shear is removed from the model (blue points). Anti-aligned shears exhibit the opposite behaviour (red points). }
    \label{Figure: shear comp mocks}
\end{figure}

The orientation of the mass distribution is robustly measured, with mean difference $\langle | \phi^\mathrm{PL+ext}_\mathrm{mass}-\phi^\mathrm{PL}_\mathrm{mass} | \rangle\sim1^\circ$ and only $\sim1^\circ$ of scatter when shear is excluded (figure~\ref{Figure: shear comp mocks}). The axis ratio is also hardly changed, with $\langle q^\mathrm{PL+ext}-q^\mathrm{PL}\rangle=0.01\pm0.05$. These values are so robust because all the MGE components share a common axis, which is therefore well-defined. 

However, the Einstein radii and slopes of the power-law mass model are systematically biased, by an amount that depends on the relative orientation of the shear and the mass (Einstein radii and power-law slopes are known to be degenerate parameters: see e.g.\ figure~5 of \citetalias{Etherington2022}). 
For lenses whose shears were aligned with the mass distribution, removing $\gamma^\mathrm{ext}$ increases the mean best-fit power-law slope by $0.20\%\pm0.04$ (blue points in figure~\ref{Figure: shear comp mocks}; their mean best-fit Einstein radii decrease by $0.2\pm0.03\%$).
For lenses whose shears were anti-aligned, the power-law slopes decrease by $0.07\pm0.03$ (red points in figure~\ref{Figure: shear comp mocks}; their mean best-fit Einstein radii increase by $0.08\pm0.04$).
Across the entire sample, Einstein radii had been correctly measured when including $\gamma^\mathrm{ext}$ \citep[within $0.05\pm0.17\%$;][]{Cao2022}, but are systematically underestimated by $0.2\%\pm0.05\%$ if shear is excluded. 

Our measurements might be caused by a bias described by \cite{Kochanek2020} in the radial structure of a lens, when a model has too few azimuthal degrees of freedom. Kochanek provides the example of fitting a power-law model to a lens whose ellipticty increases with radius: the density slope is forced to spuriously shallow values to balance the shear inside the Einstein radius relative to the shear outside it. Furthermore, \cite{VanDeVyvere2022a} found that decreasing ellipticity outside the Einstein radius spuriously increases $\gamma^\mathrm{ext}$ for a power-law model, while ellipticity gradients inside or at the Einstein radius mainly bias the power-law slope. In our mocks, the distribution of mass has an ellipticity which varies as a function of radius. We attempted to find a correlation of the measured shear properties (e.g. position angle) with the ellipticity variation of the mock mass distributions. However, we were unable to detect a statistically significant correlation. We interpret this as the measured shear depending on other properties of each lens (e.g. where the lensed source's arcs appear in the image-plane). A more detailed investigation is beyond the scope of this work.

% Our mocks whose shears align with the major axis may have a distribution of mass whose ellipticity increases with radius, and vice versa. 

\subsubsection{Real lenses}
\label{section: shear tests real}

\begin{figure}
    \centering
    \includegraphics[width=\linewidth]{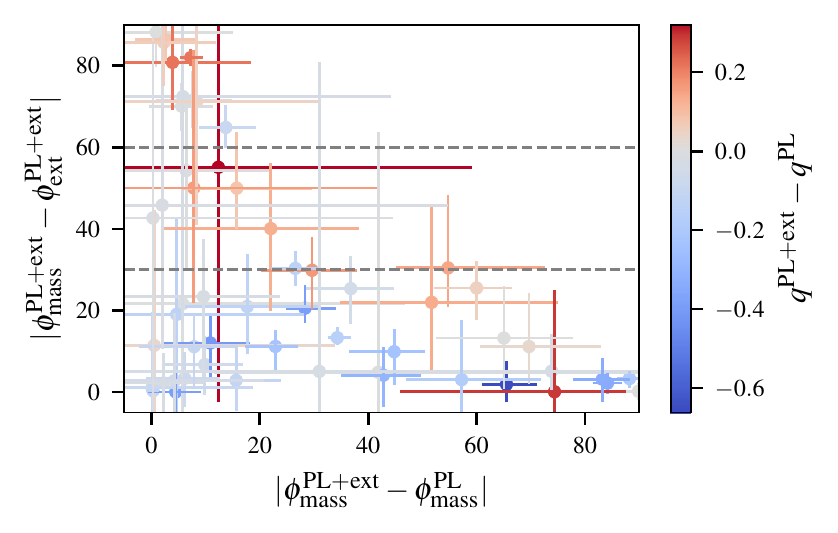}
    \caption{Orientation angle offset of the external shear from the PL+ext mass distribution ($\phi^\mathrm{PL+ext}_\mathrm{mass}-\phi^\mathrm{PL+ext}_\mathrm{ext}$) as a function of the difference in orientation angle when the mass distribution is fitted with and without an external shear ($\phi^\mathrm{PL+ext}_\mathrm{mass}-\phi^\mathrm{PL}_\mathrm{mass}$) for the observed SLACS and GALLERY  samples. Scatter points are coloured by the difference in axis ratio of the models fitted with a PL+ext and a PL ($q^\mathrm{PL+ext}-q^\mathrm{PL}$).}
    \label{Figure:shear comp data}
\end{figure}

\begin{figure*}
    \centering
    \includegraphics[width=\linewidth]{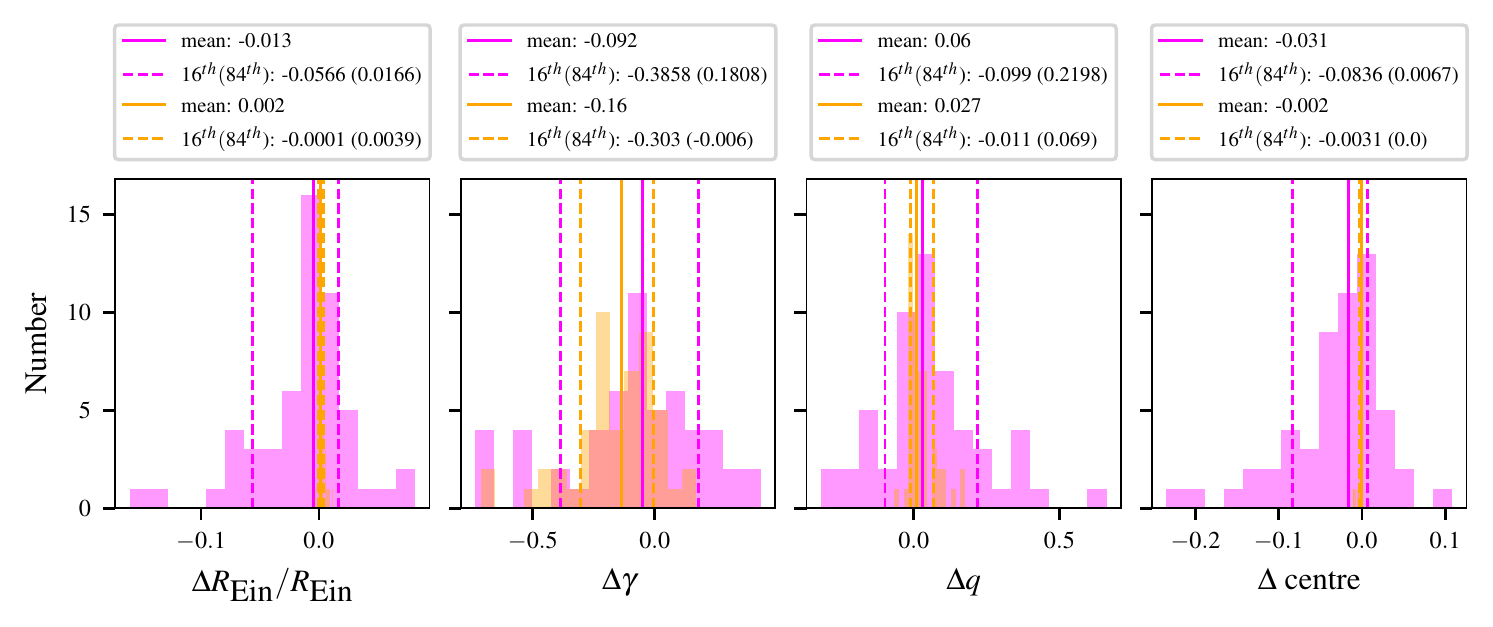}
    \caption[Histograms comparing inferred model parameters with and without shear included in the model fit for observed and mock data samples.]{Histograms of the difference between inferred PL mass distribution model parameters with external shear as free parameters, minus those without shear, for the observed (pink) and mock data samples (orange). %\rjm{The two colours really don't look that different. How about gray and black (or red)?} 
    From left to right panels the parameters are: fractional Einstein radius, logarithmic slope of density profile, axis ratio of mass distribution, and radial distance of the centre of the mass distribution, in arcseconds.}
    \label{Figure: shear comp histrograms}
\end{figure*}

\begin{figure}
    \centering
    \includegraphics[width=\linewidth]{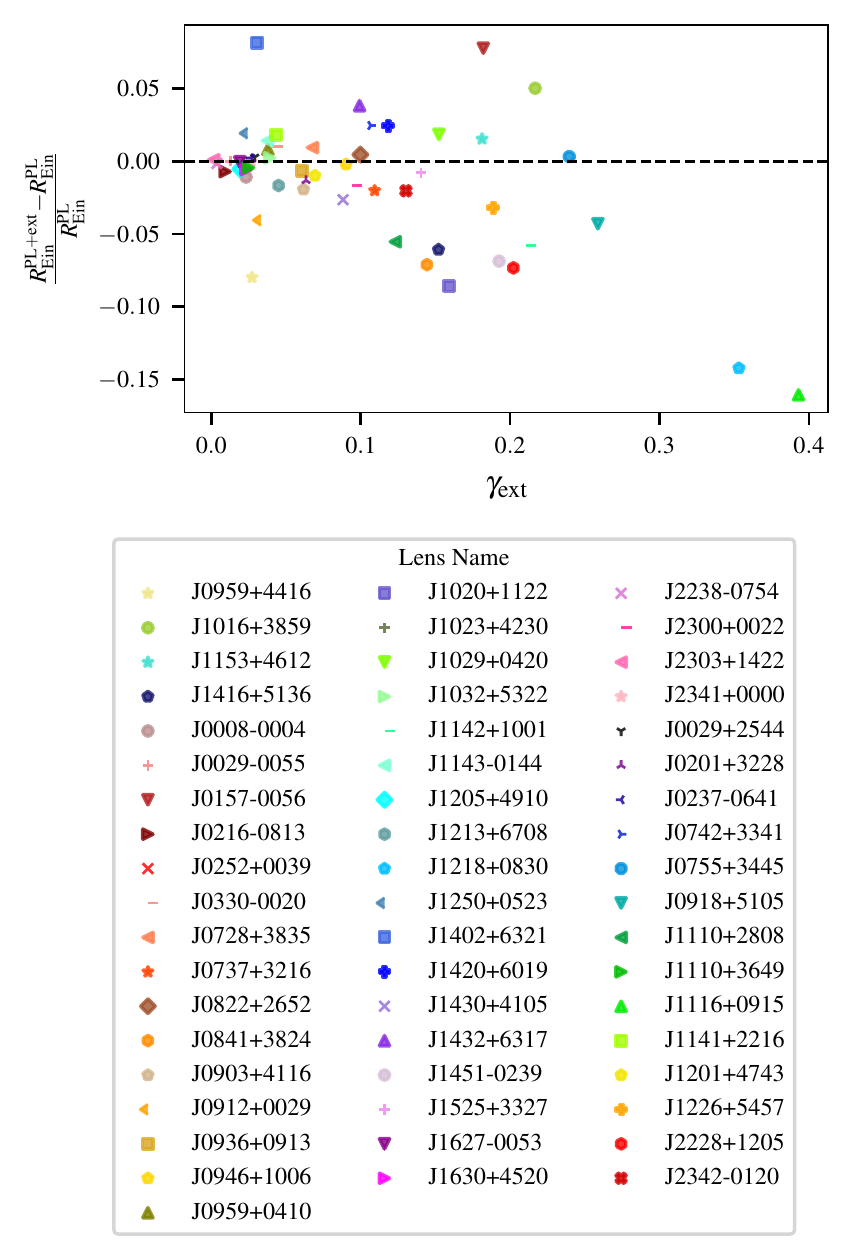}
    \caption{Fractional difference between Einstein radii inferred for models with and without the free parameters for external shear, as a function of the external shear amplitude.}
    \label{Figure: Rein v shear mag}
\end{figure}

For SLACS and GALLERY lenses, the orientation of the mass distribution changes considerably when external shear is removed, with $\langle|\phi^\mathrm{PL+ext}_\mathrm{mass}-\phi^\mathrm{PL}_\mathrm{mass}|\rangle\sim27^\circ$ and $\sim27^\circ$ scatter (figure~\ref{Figure:shear comp data}; note the different scale on the horizontal axis to figure~\ref{Figure: shear comp mocks}). The best-fit axis ratio (indicated by colour in figure~\ref{Figure:shear comp data}) also increases by $0.18\pm0.016$ for aligned systems, which become more spherical; decreases by $0.08\pm0.10$ for anti-aligned systems, which become more elliptical; and is inconsistent for the rest, with mean change $\langle q^\mathrm{PL+ext}-q^\mathrm{PL}\rangle=-0.02\pm0.04$.

Neither the Einstein radii nor slopes of the power-law mass model are systematically biased when external shear is removed. We find $\langle\Delta R_\mathrm{Ein}/R_\mathrm{Ein}\rangle=-0.013^{+0.030}_{-0.043}$, where $\Delta R_\mathrm{Ein}\equiv R_\mathrm{Ein}^\mathrm{PL+ext}-R_\mathrm{Ein}^\mathrm{PL}$. However, the best fit values scatter about the same mean, such that ${\langle\Delta R_{\mathrm Ein}^2/R_{\mathrm Ein}\rangle}=0.045^2$, which is two orders of magnitude larger than the equivalent $0.0003^2$ value for the mocks (figure~\ref{Figure: shear comp histrograms}). Furthermore, %although there does not appear to be a strong linear correlation between the fractional change in Einstein radius and the amplitude of $\gamma^{\mathrm{ext}}$, 
the scatter increases with increasing external shear magnitude (figure~\ref{Figure: Rein v shear mag}). %There is, also greater change in the best-fit power-law slopes, with $\langle\gamma^\mathrm{PL+ext}-\gamma^\mathrm{PL}\rangle=-0.09\pm0.28$. \rjm{is the error the error on the mean is it the rms?}.

The best-fit centre of the mass distribution moves on average by $-0.031\pm0.061\arcsec$ when external shear parameters are introduced, and on average remains $0.05\pm0.06\arcsec$ from the centre of light. With or without shear, this offset is an order of magnitude greater than in the mock data.

We suspect this indicates that the distribution of mass in real galaxies is more complex than that in the mocks: for example with multiple components that are rotationally offset from one another and no single, well-defined axis of ellipticity.
   
\section{Interpretation}\label{Discussion}

\subsection{Strong lensing external shears are not measuring shear}
We find that external shear, as measured by galaxy-galaxy strong lensing, does not correlate well with the true shear along a line-of-sight (as measured independently by weak lensing; Section~\ref{wl comp} and Figure~\ref{Figure: wl shear comp}). Best-fit values of external shear are also frequently several times higher than that along typical lines of sight through the Universe, so only a small fraction of it could be due to true shear. Rather, external shear tends to align with either the major axis or minor axis of the lens mass distribution (Section ~\ref{external shear behaviour}). It appears to be compensating for the inflexibility of typical mass models (here an elliptical power-law) to represent the complex distributions of mass. A substantial portion of that complexity appears to be diskiness and boxiness (Section~\ref{sec:boxiness_disciness}), especially in mock data. In real data, isophotal twists, elliptical gradients, and offsets in the centres and alignments of dark and stellar matter all increase uncertainty on model parameters. These have all been seen in the stellar mass of SLACS lenses \citep{Nightingale2019}. We have not been able to quantify how much each source contributes to the scatter or bias in a real measurement of external shear. 

\cite{Hogg2022} suggest a different, `minimal line-of-sight' way of parameterising the shear that is less degenerate with lens model parameters, but this is still subject to biases in the shear parameters when simplifying assumptions are made for the lens model. For complex distributions of lens mass, the false inference of external shear will pose a challenge for efforts to use strong lensing to measure cosmic shear \citep[e.g.][]{Birrer2017, Fleury2021}. 

\subsection{Implications for strong lensing science goals}

Including external shear alters the best-fit values of other parameters (Section~\ref{section: shear tests}). In mock data, some parameters do move closer to the known truth: the mean error in power law slopes is $-0.02\pm0.10$ with shear, or $-0.14\pm0.20$ without it. A $\sim0.2\%$ bias in Einstein radii also disappears with shear. However, the shear is not physically real and, contributing zero convergence, does not correspond to a physically meaningful distribution of mass. For studies that rely on accurate reconstructions of the mass distribution (e.g.\ galaxy evolution, dark matter physics, and the Hubble constant), this degeneracy with key parameters will eventually limit statistical precision.
For example, a key result of the SLACS survey was that elliptical galaxies have isothermal ($\gamma$=$2$) mass profiles \citep{Gavazzi2007,Vegetti2009,Auger2010}. Including external shear, \citetalias{Etherington2022a} measured $\langle\gamma\rangle=2.0756^{+0.023}_{-0.024}$ for SLACS galaxies; here we measure $\langle\gamma\rangle=2.0159=^{+0.027}_{-0.032}$ (Figure~\ref{Figure: shear comp histrograms}), highlighting the systematic uncertainty that will remain fixed even if the sample size increases. Furthermore, a PL+ext model leaves false detections of subhalos in a mock disky galaxy \citep{He2022} or real HST data \citep{Nightingale2022}.

The Hubble constant $H_0$ can be measured from the time delay between multiple images \citep{Suyu2017, Wong2019, Birrer2019}. However, assuming specific functional forms for the mass model can artificially break the mass-sheet transformation. % which, given the invariance of the product of the Hubble constant and the time delay, can lead to systematic biases. 
Our tests on mock data (Section~\ref{section: shear tests mock}) demonstrate a coupling, predicted by  \cite{Kochanek2020}, between angular and radial structure. Oversimple models bias the slope, and hence bias $H_0$. \citetalias{Cao2022} estimated $\sim$9\% bias in $H_0$ when using a PL+ext model to interpret time delays generated from gNFW+MGE lenses. The angular degrees of freedom added by `external' shear are insufficient to compensate for even this complexity. Given that our analysis of SLACS and GALLERY lenses indicate even more angular degrees of freedom, such as twists in the mass distribution, biases for real lensing systems will likely be closer to the 20--50\% suggested by other studies \citep{Schneider2013a, Xu2016, Kochanek2019, Gomer2018, Gomer2020, Gomer2021}. More flexible models, such as adding an internal mass sheet to the PL+ext model which is contstrained via stellar dynamics, have been introduced to mitigate these systematics.

Combining $H_0$ measurements from a large population of lenses might average away individual biases up to $10\mathrm{km s}^{-1}\mathrm{Mpc}^{-1}$, on the assumption that boxy and disky mass distributions are equally well represented \citep{VanDeVyvere2022a}. However, the diskiness of the mass distribution correlates with the diskiness of the (observable) PL+ext critical curves (Section~\ref{sec:crit_curve_mass_distribution}), and our observationally-selected sample contains an overpopulation of 71\% disky galaxies (Section~\ref{sec:diskiness_is_ext_shear}). More flexible mass models (e.g. a PL with an internal mass sheet) have also been introduced \citep{Birrer2020}, which infer unbiased $H_0$ values on mock lens samples \citep{Ding2021}. %Further, we suspect the alignment (and anti-alignment) of the shear with the mass profile is a result of a particular feature in the underlying mass distribution, and we find a preference for aligned shears ($70\%$ of the sample). It is plausible these sub-samples would be biased in opposite directions, that would not average out at a population level. 
Moreover, \cite{VanDeVyvere2022} did not investigate the effect of $b_4$ `twisting' perturbations (Section~\ref{section:twisting}) or mis-centering (Section~\ref{section: shear tests real}), both of which we observe in the critical curves of real galaxies, and whose effects may not average away. For example, the centre of mass in the H0LiCOW model of lens WFI\,2033-4723 is offset from the centre of light by $\sim$10$\times$ astrometric uncertainty \citep{Suyu2010,Barrera2021}, and a similar offset in iPTF16geu increases asymmetry \citep{Diego2022}.
Such offsets are unphysical \citep{schaller15}, so it is not clear that complexities in the mass distributions of real lenses can be safely ignored by averaging over a population.

\subsection{Future work: more complex mass models are needed}

External shear does not appear to be sufficient as (the sole) parameter to encode all the complexity in real lenses. Although Einstein radii are expected to be `model independent' within ${\sim}2\%$ uncertainty \cite{Bolton2008, Sonnenfeld2013b}, the 4.5\% RMS fractional difference we measure with and without shear, suggests an unknown unknown. 
%Furthermore, the centre of the mass distribution moves on average by $-0.031\pm0.061\arcsec$ when the shear parameters are introduced, but on average remains $0.05\pm0.06\arcsec$ from the centre of light \citep[c.f.][]{schaller15}. With or without shear, the offset is an order of magnitude greater in real data than in the mocks. 
A single power-law model leads to mass discrepancies with stellar dynamics \citep{Etherington2022a}, and spurious false-positives in searches for dark matter subhalos \citep{Nightingale2022}.
Further work \citep[e.g.][]{Cao2022, VanDeVyvere2020, VanDeVyvere2022a} to understand the types of asymmetries that must be accounted for in the lens modelling, and the possibility of constraining such models, will be invaluable. 

What parametric forms allow sufficient complexity --- in a minimum number of parameters that need to be constrained? Even our gNFW+MGE mocks still do not capture the full complexity of real lenses, but their MGEs were forced to be aligned. Perhaps the model could have both $a_4$ and $b_4$ perturbations, varying as a function of radius. Cluster-scale models frequently use a soft core inside some scale radius, and \citep{Limousin2022} add B-spline functions. The number of free parameters must be balanced with the information available: pixellated mass models are generally underconstrained (and although they might be able to fit observations without external shear; \citealt{Valls2006}, some line-of-sight shear is \textit{expected} for galaxy-scale lenses). To instead increase the available information, the total mass could be decomposed into dark matter and stellar components, with the latter informed by the lens light (which is otherwise a nuisance). Whatever the parametric form, the azimuthal degrees of freedom must be defined carefully to avoid the bias described by \cite{Kochanek2020} on the inference of the radial mass distribution. We suggest more studies \citep[e.g.][\citetalias{Cao2022}]{VanDeVyvere2020, VanDeVyvere2022a}, embedded deeply in each specific science case, to quantify the impact of simplifying assumptions.

%\subsection{Future work: new opportunities}

A silver lining is that the sensitivity of galaxy-galaxy strong lensing data may be a new opportunity. Both our results and those of \cite{VanDeVyvere2022} suggest that it might be possible to measure the diskiness or boxiness of galaxies' dark matter halos. Our results are even more optimistic: if measurable multipole perturbations in critical curves traces those in the distributions of both mass and light, then one could study the dark morphology of galaxies at high redshift with relative ease. 

\section*{Data Availability}

Text files and images of every model-fit performed in this work are available at \url{https://zenodo.org/record/6104823}. Full \texttt{dynesty} chains of every fit are available upon request.

\section*{Software Citations}

This work uses the following software packages:
\begin{itemize}
\item
\href{https://github.com/astropy/astropy}{{Astropy}}
\citep{astropy1, astropy2}

\item
\href{https://github.com/dfm/corner.py}{{Corner.py}}
\citep{corner}

\item
\href{https://github.com/joshspeagle/dynesty}{{Dynesty}}
\citep{dynesty}

\item
\href{https://github.com/matplotlib/matplotlib}{{Matplotlib}}
\citep{matplotlib}

\item
\href{numba` https://github.com/numba/numba}{{Numba}}
\citep{numba}

\item
\href{https://github.com/numpy/numpy}{{NumPy}}
\citep{numpy}

\item
\href{https://github.com/rhayes777/PyAutoFit}{{PyAutoFit}}
\citep{pyautofit}

\item
\href{https://github.com/Jammy2211/PyAutoGalaxy}{{PyAutoGalaxy}}
\citep{pyautogalaxy}

\item
\href{https://github.com/Jammy2211/PyAutoLens}{{PyAutoLens}}
\citep{Nightingale2015, Nightingale2018, Nightingale2021}

\item
\href{https://www.python.org/}{{Python}}
\citep{python}

\item
\href{https://github.com/scikit-image/scikit-image}{{Scikit-image}}
\citep{scikit-image}

\item
\href{https://github.com/scikit-learn/scikit-learn}{{Scikit-learn}}
\citep{scikit-learn}

\item
\href{https://github.com/scipy/scipy}{{Scipy}}
\citep{scipy}

\item
\href{https://www.sqlite.org/index.html}{{SQLite}}
\citep{sqlite}

\end{itemize}

\section*{Acknowledgements}

We thank Yiping Shu, Guillaume Desprez, Johan Richard, Mathilde Jauzac, and Keichi Umetsu for providing data from previously published papers, and Liliya Williams and Adi Zitrin for stimulating discussions.

AE is supported by UK STFC via grants ST/R504725/1 and ST/T506047/1. JN and RM are supported by STFC via grant ST/T002565/1, and the UK Space Agency via grant ST/W002612/1. XYC and RL are supported by the National Nature Science Foundation of China (via grants 11988101, 11773032, 12022306), the China Manned Space Project (science research grants CMS-CSST-2021-B01, CMS-CSST-2021-A01) and by the K.C.Wong Education Foundation. QH, AA, SMC, and CSF are supported by %the European Research Council (ERC) through Advanced Investigator grant DMIDAS (GA~786910). 
the ERC via grant GA~786910. MJ is supported by UKRI via grant MR/S017216/1. This work used both the Cambridge Service for Data Driven Discovery (CSD3) and the DiRAC Data-Centric system, which are operated by the University of Cambridge and Durham University on behalf of the STFC DiRAC HPC Facility (www.dirac.ac.uk). These were funded by BIS capital grant ST/K00042X/1, STFC capital grants ST/P002307/1, ST/R002452/1, ST/H008519/1, ST/K00087X/1, STFC Operations grants ST/K003267/1, ST/K003267/1, and Durham University. DiRAC is part of the UK National E-Infrastructure.

\bibliographystyle{mnras}
\bibliography{references2.bib, references.bib, software.bib} % if your bibtex file is called example.bib

\begin{thebibliography}{}
\makeatletter
\relax
\def\mn@urlcharsother{\let\do\@makeother \do\$\do\&\do\#\do\^\do\_\do\%\do\~}
\def\mn@doi{\begingroup\mn@urlcharsother \@ifnextchar [ {\mn@doi@}
  {\mn@doi@[]}}
\def\mn@doi@[#1]#2{\def\@tempa{#1}\ifx\@tempa\@empty \href
  {http://dx.doi.org/#2} {doi:#2}\else \href {http://dx.doi.org/#2} {#1}\fi
  \endgroup}
\def\mn@eprint#1#2{\mn@eprint@#1:#2::\@nil}
\def\mn@eprint@arXiv#1{\href {http://arxiv.org/abs/#1} {{\tt arXiv:#1}}}
\def\mn@eprint@dblp#1{\href {http://dblp.uni-trier.de/rec/bibtex/#1.xml}
  {dblp:#1}}
\def\mn@eprint@#1:#2:#3:#4\@nil{\def\@tempa {#1}\def\@tempb {#2}\def\@tempc
  {#3}\ifx \@tempc \@empty \let \@tempc \@tempb \let \@tempb \@tempa \fi \ifx
  \@tempb \@empty \def\@tempb {arXiv}\fi \@ifundefined
  {mn@eprint@\@tempb}{\@tempb:\@tempc}{\expandafter \expandafter \csname
  mn@eprint@\@tempb\endcsname \expandafter{\@tempc}}}

\bibitem[\protect\citeauthoryear{Ade et~al.,}{Ade et~al.}{2016}]{Ade2016}
Ade P.~A.,  et~al., 2016, \mn@doi [Astronomy and Astrophysics]
  {10.1051/0004-6361/201525830}, 594

\bibitem[\protect\citeauthoryear{Amorisco et~al.,}{Amorisco
  et~al.}{2022}]{Amorisco2021}
Amorisco N.~C.,  et~al., 2022, \mn@doi [Monthly Notices of the Royal
  Astronomical Society] {10.1093/mnras/stab3527}, 510, 2464

\bibitem[\protect\citeauthoryear{{Astropy Collaboration}}{{Astropy
  Collaboration}}{2013}]{astropy1}
{Astropy Collaboration} 2013, \mn@doi [A\&A] {10.1051/0004-6361/201322068},
  \href {http://adsabs.harvard.edu/abs/2013A%26A...558A..33A} {558, A33}

\bibitem[\protect\citeauthoryear{Auger, Treu, Bolton, Gavazzi, Koopmans,
  Marshall, Moustakas  \& Burles}{Auger et~al.}{2010}]{Auger2010}
Auger M.~W.,  Treu T.,  Bolton A.~S.,  Gavazzi R.,  Koopmans L.~V.,  Marshall
  P.~J.,  Moustakas L.~A.,   Burles S.,  2010, \mn@doi [Astrophysical Journal]
  {10.1088/0004-637X/724/1/511}, 724, 511

\bibitem[\protect\citeauthoryear{{Barrera}, {Williams}, {Coles}  \&
  {Denzel}}{{Barrera} et~al.}{2021}]{Barrera2021}
{Barrera} B.,  {Williams} L. L.~R.,  {Coles} J.~P.,   {Denzel} P.,  2021,
  \mn@doi [The Open Journal of Astrophysics] {10.21105/astro.2108.04348}, \href
  {https://ui.adsabs.harvard.edu/abs/2021OJAp....4E..12B} {4, 12}

\bibitem[\protect\citeauthoryear{{Bertin} \& {Arnouts}}{{Bertin} \&
  {Arnouts}}{1996}]{sextractor}
{Bertin} E.,  {Arnouts} S.,  1996, \mn@doi [Astron. Astrophys. Suppl. Ser.]
  {10.1051/aas:1996164}, 117, 393

\bibitem[\protect\citeauthoryear{Birrer, Welschen, Amara  \& Refregier}{Birrer
  et~al.}{2017}]{Birrer2017}
Birrer S.,  Welschen C.,  Amara A.,   Refregier A.,  2017, \mn@doi [Journal of
  Cosmology and Astroparticle Physics] {10.1088/1475-7516/2017/04/049}, 2017

\bibitem[\protect\citeauthoryear{Birrer et~al.,}{Birrer
  et~al.}{2019}]{Birrer2019}
Birrer S.,  et~al., 2019, \mn@doi [Monthly Notices of the Royal Astronomical
  Society] {10.1093/mnras/stz200}, 484, 4726

\bibitem[\protect\citeauthoryear{Birrer et~al.,}{Birrer
  et~al.}{2020}]{Birrer2020}
Birrer S.,  et~al., 2020, \mn@doi [Astronomy {\&} Astrophysics]
  {10.1051/0004-6361/202038861}, 1

\bibitem[\protect\citeauthoryear{Bolton, Burles, Koopmans, Treu, Gavazzi,
  Moustakas, Wayth  \& Schlegel}{Bolton et~al.}{2008a}]{Bolton2008a}
Bolton A.~S.,  Burles S.,  Koopmans L. V.~E.,  Treu T.,  Gavazzi R.,  Moustakas
  L.~A.,  Wayth R.,   Schlegel D.~J.,  2008a, \mn@doi [The Astrophysical
  Journal] {10.1086/589327}

\bibitem[\protect\citeauthoryear{Bolton, Treu, Koopmans, Gavazzi, Moustakas,
  Burles, Schlegel  \& Wayth}{Bolton et~al.}{2008b}]{Bolton2008}
Bolton A.~S.,  Treu T.,  Koopmans L. V.~E.,  Gavazzi R.,  Moustakas L.~A.,
  Burles S.,  Schlegel D.~J.,   Wayth R.,  2008b, \mn@doi [The Astrophysical
  Journal] {10.1086/589989}, 684, 248

\bibitem[\protect\citeauthoryear{Bolton et~al.,}{Bolton
  et~al.}{2012}]{Bolton2012}
Bolton A.~S.,  et~al., 2012, \mn@doi [Astrophysical Journal]
  {10.1088/0004-637X/757/1/82}, 757

\bibitem[\protect\citeauthoryear{Cao et~al.,}{Cao et~al.}{2022}]{Cao2022}
Cao X.,  et~al., 2022, \mn@doi [Research in Astronomy and Astrophysics]
  {10.1088/1674-4527/ac3f2b}, 22

\bibitem[\protect\citeauthoryear{Cappellari}{Cappellari}{2002}]{Cappellari2002}
Cappellari M.,  2002, \mn@doi [Monthly Notices of the Royal Astronomical
  Society] {10.1046/j.1365-8711.2002.05412.x}, 333, 400

\bibitem[\protect\citeauthoryear{Cappellari et~al.,}{Cappellari
  et~al.}{2013}]{Cappellari2013}
Cappellari M.,  et~al., 2013, \mn@doi [Monthly Notices of the Royal
  Astronomical Society] {10.1093/mnras/stt562}, 432, 1709

\bibitem[\protect\citeauthoryear{Despali, Lovell, Vegetti, Crain  \&
  Oppenheimer}{Despali et~al.}{2019}]{Despali2019}
Despali G.,  Lovell M.,  Vegetti S.,  Crain R.~A.,   Oppenheimer B.~D.,  2019,
  \mn@doi [Monthly Notices of the Royal Astronomical Society]
  {10.1093/mnras/stz3068}, 17, 1

\bibitem[\protect\citeauthoryear{{Desprez}, {Richard}, {Jauzac}, {Martinez},
  {Siana}  \& {Cl{\'e}ment}}{{Desprez} et~al.}{2018}]{Desprez2018}
{Desprez} G.,  {Richard} J.,  {Jauzac} M.,  {Martinez} J.,  {Siana} B.,
  {Cl{\'e}ment} B.,  2018, \mn@doi [\mnras] {10.1093/mnras/sty1666}, \href
  {https://ui.adsabs.harvard.edu/abs/2018MNRAS.479.2630D} {479, 2630}

\bibitem[\protect\citeauthoryear{{Diego}, {Bernstein}, {Chen}, {Goobar},
  {Johansson}, {Kelly}, {M{\"o}rtsell}  \& {Nightingale}}{{Diego}
  et~al.}{2022}]{Diego2022}
{Diego} J.~M.,  {Bernstein} G.,  {Chen} W.,  {Goobar} A.,  {Johansson} J.~P.,
  {Kelly} P.~L.,  {M{\"o}rtsell} E.,   {Nightingale} J.~W.,  2022, \mn@doi
  [\aap] {10.1051/0004-6361/202143009}, \href
  {https://ui.adsabs.harvard.edu/abs/2022A&A...662A..34D} {662, A34}

\bibitem[\protect\citeauthoryear{Ding et~al.,}{Ding et~al.}{2021}]{Ding2021}
Ding X.,  et~al., 2021, \mn@doi [Monthly Notices of the Royal Astronomical
  Society] {10.1093/mnras/stab484}, 503, 1096

\bibitem[\protect\citeauthoryear{Etherington et~al.,}{Etherington
  et~al.}{2022}]{Etherington2022}
Etherington A.,  et~al., 2022, \mn@doi [MNRAS] {10.1093/mnras/stac2639}, 517,
  3275

\bibitem[\protect\citeauthoryear{Etherington et~al.,}{Etherington
  et~al.}{2023}]{Etherington2022a}
Etherington A.,  et~al., 2023, MNRAS, 521, 6005

\bibitem[\protect\citeauthoryear{{Faure} et~al.,}{{Faure}
  et~al.}{2008}]{Faure2008}
{Faure} C.,  et~al., 2008, \mn@doi [\apjs] {10.1086/526426}, \href
  {https://ui.adsabs.harvard.edu/abs/2008ApJS..176...19F} {176, 19}

\bibitem[\protect\citeauthoryear{{Fleury}, {Larena}  \& {Uzan}}{{Fleury}
  et~al.}{2021}]{Fleury2021}
{Fleury} P.,  {Larena} J.,   {Uzan} J.-P.,  2021, \mn@doi [JCAP]
  {10.1088/1475-7516/2021/08/024}, \href
  {https://ui.adsabs.harvard.edu/abs/2021JCAP...08..024F} {2021, 024}

\bibitem[\protect\citeauthoryear{Foreman-Mackey}{Foreman-Mackey}{2016}]{corner}
Foreman-Mackey D.,  2016, \mn@doi [The J. Open Source Softw.]
  {10.21105/joss.00024}, 1, 24

\bibitem[\protect\citeauthoryear{{Gavazzi}, {Treu}, {Rhodes}, {Koopmans},
  {Bolton}, {Burles}, {Massey}  \& {Moustakas}}{{Gavazzi}
  et~al.}{2007}]{Gavazzi2007}
{Gavazzi} R.,  {Treu} T.,  {Rhodes} J.~D.,  {Koopmans} L. V.~E.,  {Bolton}
  A.~S.,  {Burles} S.,  {Massey} R.~J.,   {Moustakas} L.~A.,  2007, \mn@doi
  [\apj] {10.1086/519237}, \href
  {https://ui.adsabs.harvard.edu/abs/2007ApJ...667..176G} {667, 176}

\bibitem[\protect\citeauthoryear{{Gomer} \& {Williams}}{{Gomer} \&
  {Williams}}{2018}]{Gomer2018}
{Gomer} M.~R.,  {Williams} L. L.~R.,  2018, \mn@doi [\mnras]
  {10.1093/mnras/stx3294}, \href
  {https://ui.adsabs.harvard.edu/abs/2018MNRAS.475.1987G} {475, 1987}

\bibitem[\protect\citeauthoryear{{Gomer} \& {Williams}}{{Gomer} \&
  {Williams}}{2020}]{Gomer2020}
{Gomer} M.,  {Williams} L.~L.~R.,  2020, \mn@doi [JCAP]
  {10.1088/1475-7516/2020/11/045}, \href
  {https://ui.adsabs.harvard.edu/abs/2020JCAP...11..045G} {2020, 045}

\bibitem[\protect\citeauthoryear{{Gomer} \& {Williams}}{{Gomer} \&
  {Williams}}{2021}]{Gomer2021}
{Gomer} M.~R.,  {Williams} L. L.~R.,  2021, \mn@doi [\mnras]
  {10.1093/mnras/stab930}, \href
  {https://ui.adsabs.harvard.edu/abs/2021MNRAS.504.1340G} {504, 1340}

\bibitem[\protect\citeauthoryear{Graham \& Driver}{Graham \&
  Driver}{2005}]{Graham2005}
Graham A.~W.,  Driver S.~P.,  2005, \mn@doi [Publications of the Astronomical
  Society of Australia] {10.1071/AS05001}, 22, 118

\bibitem[\protect\citeauthoryear{Harvey}{Harvey}{2020}]{Harvey2020b}
Harvey D.,  2020, \mn@doi [Monthly Notices of the Royal Astronomical Society]
  {10.1093/mnras/staa2522}, 498, 2871

\bibitem[\protect\citeauthoryear{{Harvey}, {Tam}, {Jauzac}, {Massey}  \&
  {Rhodes}}{{Harvey} et~al.}{2019}]{Harvey2019}
{Harvey} D.,  {Tam} S.-I.,  {Jauzac} M.,  {Massey} R.,   {Rhodes} J.,  2019,
  arXiv e-prints, \href {https://ui.adsabs.harvard.edu/abs/2019arXiv191106333H}
  {p. arXiv:1911.06333}

\bibitem[\protect\citeauthoryear{He et~al.,}{He et~al.}{2022a}]{HeAmy2022a}
He Q.,  et~al., 2022a, \mn@doi [MNRAS] {10.1093/mnras/stac191}, 511, 3046

\bibitem[\protect\citeauthoryear{He et~al.,}{He et~al.}{2022b}]{HeAmy2022b}
He Q.,  et~al., 2022b, \mn@doi [MNRAS] {10.1093/mnras/stac759}, 512, 5862

\bibitem[\protect\citeauthoryear{He et~al.,}{He et~al.}{2023}]{He2022}
He Q.,  et~al., 2023, \mn@doi [MNRAS] {10.1093/mnras/stac2779}, 518, 220

\bibitem[\protect\citeauthoryear{Hezaveh et~al.,}{Hezaveh
  et~al.}{2016}]{Hezaveh2016}
Hezaveh Y.~D.,  et~al., 2016, \mn@doi [The Astrophysical Journal]
  {10.3847/0004-637x/823/1/37}, 823, 37

\bibitem[\protect\citeauthoryear{{Hilbert}, {White}, {Hartlap}  \&
  {Schneider}}{{Hilbert} et~al.}{2007}]{Hilbert2007}
{Hilbert} S.,  {White} S. D.~M.,  {Hartlap} J.,   {Schneider} P.,  2007,
  \mn@doi [\mnras] {10.1111/j.1365-2966.2007.12391.x}, \href
  {https://ui.adsabs.harvard.edu/abs/2007MNRAS.382..121H} {382, 121}

\bibitem[\protect\citeauthoryear{Hipp}{Hipp}{2020}]{sqlite}
Hipp R.~D.,  2020, {SQLite}, \url {https://www.sqlite.org/index.html}

\bibitem[\protect\citeauthoryear{Hogg, Fleury, Larena  \& Martinelli}{Hogg
  et~al.}{2022}]{Hogg2022}
Hogg N.~B.,  Fleury P.,  Larena J.,   Martinelli M.,  2022, 19, 1

\bibitem[\protect\citeauthoryear{Hunter}{Hunter}{2007}]{matplotlib}
Hunter J.~D.,  2007, \mn@doi [Comput Sci Eng] {10.1109/MCSE.2007.55}, 9, 90

\bibitem[\protect\citeauthoryear{Keeton, Kochanek  \& Seljak}{Keeton
  et~al.}{1997}]{Keeton1997}
Keeton C.~R.,  Kochanek C.~S.,   Seljak U.,  1997, \mn@doi [The Astrophysical
  Journal] {10.1086/304172}, 482, 604

\bibitem[\protect\citeauthoryear{Kilbinger}{Kilbinger}{2015}]{Kilbinger2015}
Kilbinger M.,  2015, \mn@doi [Reports on Progress in Physics]
  {10.1088/0034-4885/78/8/086901}, 78

\bibitem[\protect\citeauthoryear{{Kochanek}}{{Kochanek}}{2020}]{Kochanek2019}
{Kochanek} C.~S.,  2020, \mn@doi [\mnras] {10.1093/mnras/staa344}, \href
  {https://ui.adsabs.harvard.edu/abs/2020MNRAS.493.1725K} {493, 1725}

\bibitem[\protect\citeauthoryear{{Kochanek}}{{Kochanek}}{2021}]{Kochanek2020}
{Kochanek} C.~S.,  2021, \mn@doi [\mnras] {10.1093/mnras/staa4033}, \href
  {https://ui.adsabs.harvard.edu/abs/2021MNRAS.501.5021K} {501, 5021}

\bibitem[\protect\citeauthoryear{{Kochanek}, {Schneider}  \&
  {Wambsganss}}{{Kochanek} et~al.}{2004}]{saasfee}
{Kochanek} C.,  {Schneider} P.,   {Wambsganss} J.,  2004, Part 2 of
  Gravitational Lensing: Strong, Weak \& Micro, Proceedings of the 33rd
  Saas-Fee Advanced Course, \url {https://arxiv.org/abs/astro-ph/0407232}

\bibitem[\protect\citeauthoryear{Koopmans et~al.,}{Koopmans
  et~al.}{2009}]{Koopmans2009}
Koopmans L.~V.,  et~al., 2009, \mn@doi [Astrophysical Journal]
  {10.1088/0004-637X/703/1/L51}, 703

\bibitem[\protect\citeauthoryear{Kuhn, Bruderer, Birrer, Amara  \&
  R{\'{e}}fr{\'{e}}gier}{Kuhn et~al.}{2020}]{Kuhn2020}
Kuhn F.~A.,  Bruderer C.,  Birrer S.,  Amara A.,   R{\'{e}}fr{\'{e}}gier A.,
  2020, pp 0--26

\bibitem[\protect\citeauthoryear{Lam, Pitrou  \& Seibert}{Lam
  et~al.}{2015}]{numba}
Lam S.~K.,  Pitrou A.,   Seibert S.,  2015, \mn@doi [Proceedings of the Second
  Workshop on the LLVM Compiler Infrastructure in HPC - LLVM '15]
  {10.1145/2833157.2833162}, pp~1--6

\bibitem[\protect\citeauthoryear{Leauthaud et~al.,}{Leauthaud
  et~al.}{2007}]{Leauthaud2007a}
Leauthaud A.,  et~al., 2007, \mn@doi [The Astrophysical Journal Supplement
  Series] {10.1086/516598}, 172, 219

\bibitem[\protect\citeauthoryear{Li, Frenk, Cole, Gao, Bose  \& Hellwing}{Li
  et~al.}{2016}]{Li2016}
Li R.,  Frenk C.~S.,  Cole S.,  Gao L.,  Bose S.,   Hellwing W.~A.,  2016,
  \mn@doi [Monthly Notices of the Royal Astronomical Society]
  {10.1093/mnras/stw939}, 460, 363

\bibitem[\protect\citeauthoryear{Li, Frenk, Cole, Wang  \& Gao}{Li
  et~al.}{2017}]{Li2017}
Li R.,  Frenk C.~S.,  Cole S.,  Wang Q.,   Gao L.,  2017, \mn@doi [Monthly
  Notices of the Royal Astronomical Society] {10.1093/mnras/stx554}, 468, 1426

\bibitem[\protect\citeauthoryear{Li et~al.,}{Li et~al.}{2019}]{Li2019}
Li R.,  et~al., 2019, \mn@doi [Monthly Notices of the Royal Astronomical
  Society] {10.1093/mnras/stz2565}, 490, 2124

\bibitem[\protect\citeauthoryear{{Limousin}, {Beauchesne}  \&
  {Jullo}}{{Limousin} et~al.}{2022}]{Limousin2022}
{Limousin} M.,  {Beauchesne} B.,   {Jullo} E.,  2022, \mn@doi [\aap]
  {10.1051/0004-6361/202243278}, \href
  {https://ui.adsabs.harvard.edu/abs/2022A&A...664A..90L} {664, A90}

\bibitem[\protect\citeauthoryear{{Massey} et~al.,}{{Massey}
  et~al.}{2007}]{Massey2007c}
{Massey} R.,  et~al., 2007, \mn@doi [\apjs] {10.1086/516599}, \href
  {https://ui.adsabs.harvard.edu/abs/2007ApJS..172..239M} {172, 239}

\bibitem[\protect\citeauthoryear{Massey, Kitching  \& Richard}{Massey
  et~al.}{2010}]{Massey2010}
Massey R.,  Kitching T.,   Richard J.,  2010, \mn@doi [Reports on Progress in
  Physics] {10.1088/0034-4885/73/8/086901}, 73

\bibitem[\protect\citeauthoryear{{Narayan} \& {Bartelmann}}{{Narayan} \&
  {Bartelmann}}{1996}]{NarayanBartelmann1996}
{Narayan} R.,  {Bartelmann} M.,  1996, \mn@doi [arXiv e-prints]
  {10.48550/arXiv.astro-ph/9606001}, \href
  {https://ui.adsabs.harvard.edu/abs/1996astro.ph..6001N} {pp
  astro--ph/9606001}

\bibitem[\protect\citeauthoryear{{Natarajan} \& {Springel}}{{Natarajan} \&
  {Springel}}{2004}]{Natarajan2004}
{Natarajan} P.,  {Springel} V.,  2004, \mn@doi [\apjl] {10.1086/427079}, \href
  {https://ui.adsabs.harvard.edu/abs/2004ApJ...617L..13N} {617, L13}

\bibitem[\protect\citeauthoryear{Navarro, Frenk  \& White}{Navarro
  et~al.}{1997}]{Navarro1997}
Navarro J.~F.,  Frenk C.~S.,   White S. D.~M.,  1997, \mn@doi [ApJ]
  {10.1086/304888}, 490, 493

\bibitem[\protect\citeauthoryear{Nightingale \& Dye}{Nightingale \&
  Dye}{2015}]{Nightingale2015}
Nightingale J.~W.,  Dye S.,  2015, \mn@doi [Monthly Notices of the Royal
  Astronomical Society] {10.1093/mnras/stv1455}, 452, 2940

\bibitem[\protect\citeauthoryear{Nightingale, Dye  \& Massey}{Nightingale
  et~al.}{2018}]{Nightingale2018}
Nightingale J.,  Dye S.,   Massey R.,  2018, \mn@doi [Monthly Notices of the
  Royal Astronomical Society] {10.1093/mnras/sty1264}, 47, 1

\bibitem[\protect\citeauthoryear{Nightingale, Massey, Harvey, Cooper,
  Etherington, Tam  \& Hayes}{Nightingale et~al.}{2019}]{Nightingale2019}
Nightingale J.~W.,  Massey R.~J.,  Harvey D.~R.,  Cooper A.~P.,  Etherington
  A.,  Tam S.-I.,   Hayes R.~G.,  2019, \mn@doi [Monthly Notices of the Royal
  Astronomical Society] {10.1093/mnras/stz2220}

\bibitem[\protect\citeauthoryear{Nightingale, Hayes  \& Griffiths}{Nightingale
  et~al.}{2021a}]{pyautofit}
Nightingale J.~W.,  Hayes R.~G.,   Griffiths M.,  2021a, \mn@doi [J. Open
  Source Softw.] {10.21105/joss.02550}, 6, 2550

\bibitem[\protect\citeauthoryear{Nightingale et~al.,}{Nightingale
  et~al.}{2021b}]{Nightingale2021}
Nightingale J.,  et~al., 2021b, \mn@doi [Journal of Open Source Software]
  {10.21105/joss.02825}, 6, 2825

\bibitem[\protect\citeauthoryear{{Nightingale} et~al.,}{{Nightingale}
  et~al.}{2023a}]{Nightingale2022}
{Nightingale} J.~W.,  et~al., 2023a, arXiv e-prints, \href
  {https://ui.adsabs.harvard.edu/abs/2022arXiv220910566N} {p. arXiv:2209.10566}

\bibitem[\protect\citeauthoryear{Nightingale et~al.,}{Nightingale
  et~al.}{2023b}]{pyautogalaxy}
Nightingale J.~W.,  et~al., 2023b, \mn@doi [J. Open Source Softw.]
  {10.21105/joss.04475}, 8, 4475

\bibitem[\protect\citeauthoryear{{Nightingale} et~al.,}{{Nightingale}
  et~al.}{2023c}]{Nightingale2023}
{Nightingale} J.~W.,  et~al., 2023c, \mn@doi [\mnras] {10.1093/mnras/stad587},
  \href {https://ui.adsabs.harvard.edu/abs/2023MNRAS.521.3298N} {521, 3298}

\bibitem[\protect\citeauthoryear{Oldham \& Auger}{Oldham \&
  Auger}{2018}]{Oldham2018}
Oldham L.~J.,  Auger M.~W.,  2018, \mn@doi [Monthly Notices of the Royal
  Astronomical Society] {10.1093/mnras/sty065}, 476, 133

\bibitem[\protect\citeauthoryear{Pedregosa et~al.,}{Pedregosa
  et~al.}{2011}]{scikit-learn}
Pedregosa F.,  et~al., 2011, Journal of Machine Learning Research, 12, 2825

\bibitem[\protect\citeauthoryear{{Price-Whelan} et~al.,}{{Price-Whelan}
  et~al.}{2018}]{astropy2}
{Price-Whelan} A.~M.,  et~al., 2018, \mn@doi [AJ] {10.3847/1538-3881/aabc4f},
  \href {https://ui.adsabs.harvard.edu/#abs/2018AJ....156..123T} {156, 123}

\bibitem[\protect\citeauthoryear{Rhodes, Refregier  \& Groth}{Rhodes
  et~al.}{2000}]{Rhodes2000}
Rhodes J.,  Refregier A.,   Groth E.~J.,  2000, \mn@doi [The Astrophysical
  Journal] {10.1086/308902}, 536, 79

\bibitem[\protect\citeauthoryear{Ritondale, Vegetti, Despali, Auger, Koopmans
  \& McKean}{Ritondale et~al.}{2019}]{Ritondale2019}
Ritondale E.,  Vegetti S.,  Despali G.,  Auger M.~W.,  Koopmans L.~V.,   McKean
  J.~P.,  2019, \mn@doi [Monthly Notices of the Royal Astronomical Society]
  {10.1093/mnras/stz464}, 485, 2179

\bibitem[\protect\citeauthoryear{{Robertson}, {Smith}, {Massey}, {Eke},
  {Jauzac}, {Bianconi}  \& {Ryczanowski}}{{Robertson}
  et~al.}{2020}]{Robertson2020b}
{Robertson} A.,  {Smith} G.~P.,  {Massey} R.,  {Eke} V.,  {Jauzac} M.,
  {Bianconi} M.,   {Ryczanowski} D.,  2020, \mn@doi [\mnras]
  {10.1093/mnras/staa1429}, \href
  {https://ui.adsabs.harvard.edu/abs/2020MNRAS.495.3727R} {495, 3727}

\bibitem[\protect\citeauthoryear{{Schaller}, {Robertson}, {Massey}, {Bower}  \&
  {Eke}}{{Schaller} et~al.}{2015}]{schaller15}
{Schaller} M.,  {Robertson} A.,  {Massey} R.,  {Bower} R.~G.,   {Eke} V.~R.,
  2015, \mn@doi [\mnras] {10.1093/mnrasl/slv104}, \href
  {https://ui.adsabs.harvard.edu/abs/2015MNRAS.453L..58S} {453, L58}

\bibitem[\protect\citeauthoryear{Schneider \& Sluse}{Schneider \&
  Sluse}{2013}]{Schneider2013a}
Schneider P.,  Sluse D.,  2013, \mn@doi [Astronomy and Astrophysics]
  {10.1051/0004-6361/201321882}, 559, 1

\bibitem[\protect\citeauthoryear{Shu et~al.,}{Shu et~al.}{2015}]{Shu2015}
Shu Y.,  et~al., 2015, \mn@doi [Astrophysical Journal]
  {10.1088/0004-637X/803/2/71}, 803, 1

\bibitem[\protect\citeauthoryear{Shu, Bolton, Moustakas, Stern, Dey,
  Brownstein, Burles  \& Spinrad}{Shu et~al.}{2016a}]{Shu2016c}
Shu Y.,  Bolton A.~S.,  Moustakas L.~A.,  Stern D.,  Dey A.,  Brownstein J.~R.,
   Burles S.,   Spinrad H.,  2016a, \mn@doi [The Astrophysical Journal]
  {10.3847/0004-637x/820/1/43}, 820, 43

\bibitem[\protect\citeauthoryear{Shu et~al.,}{Shu et~al.}{2016b}]{Shu2016b}
Shu Y.,  et~al., 2016b, \mn@doi [The Astrophysical Journal]
  {10.3847/0004-637x/824/2/86}, 824, 86

\bibitem[\protect\citeauthoryear{{Smail}, {Ellis}  \& {Fitchett}}{{Smail}
  et~al.}{1994}]{Smail1994}
{Smail} I.,  {Ellis} R.~S.,   {Fitchett} M.~J.,  1994, \mn@doi [\mnras]
  {10.1093/mnras/270.2.245}, \href
  {https://ui.adsabs.harvard.edu/abs/1994MNRAS.270..245S} {270, 245}

\bibitem[\protect\citeauthoryear{Sonnenfeld, Treu, Gavazzi, Marshall, Auger,
  Suyu, Koopmans  \& Bolton}{Sonnenfeld et~al.}{2012}]{Sonnenfeld2012}
Sonnenfeld A.,  Treu T.,  Gavazzi R.,  Marshall P.~J.,  Auger M.~W.,  Suyu
  S.~H.,  Koopmans L.~V.,   Bolton A.~S.,  2012, \mn@doi [Astrophysical
  Journal] {10.1088/0004-637X/752/2/163}, 752

\bibitem[\protect\citeauthoryear{Sonnenfeld, Treu, Gavazzi, Suyu, Marshall,
  Auger  \& Nipoti}{Sonnenfeld et~al.}{2013}]{Sonnenfeld2013b}
Sonnenfeld A.,  Treu T.,  Gavazzi R.,  Suyu S.~H.,  Marshall P.~J.,  Auger
  M.~W.,   Nipoti C.,  2013, \mn@doi [Astrophysical Journal]
  {10.1088/0004-637X/777/2/98}, 777

\bibitem[\protect\citeauthoryear{Speagle}{Speagle}{2020}]{dynesty}
Speagle J.~S.,  2020, \mn@doi [MNRAS] {10.1093/mnras/staa278}, 493, 3132

\bibitem[\protect\citeauthoryear{Suyu}{Suyu}{2012}]{Suyu2012}
Suyu S.~H.,  2012, \mn@doi [Monthly Notices of the Royal Astronomical Society]
  {10.1111/j.1365-2966.2012.21661.x}, 426, 868

\bibitem[\protect\citeauthoryear{{Suyu}, {Marshall}, {Auger}, {Hilbert},
  {Blandford}, {Koopmans}, {Fassnacht}  \& {Treu}}{{Suyu}
  et~al.}{2010}]{Suyu2010}
{Suyu} S.~H.,  {Marshall} P.~J.,  {Auger} M.~W.,  {Hilbert} S.,  {Blandford}
  R.~D.,  {Koopmans} L.~V.~E.,  {Fassnacht} C.~D.,   {Treu} T.,  2010, \mn@doi
  [\apj] {10.1088/0004-637X/711/1/201}, \href
  {https://ui.adsabs.harvard.edu/abs/2010ApJ...711..201S} {711, 201}

\bibitem[\protect\citeauthoryear{Suyu et~al.,}{Suyu et~al.}{2017}]{Suyu2017}
Suyu S.~H.,  et~al., 2017, \mn@doi [Monthly Notices of the Royal Astronomical
  Society] {10.1093/mnras/stx483}, 468, 2590

\bibitem[\protect\citeauthoryear{Tam et~al.,}{Tam et~al.}{2020}]{Tam2020}
Tam S.~I.,  et~al., 2020, \mn@doi [Monthly Notices of the Royal Astronomical
  Society] {10.1093/MNRAS/STAA1828}, 496, 4032

\bibitem[\protect\citeauthoryear{Treu, Gavazzi, Gorecki, Marshall, Koopmans,
  Bolton, Moustakas  \& Burles}{Treu et~al.}{2009}]{Treu2009}
Treu T.,  Gavazzi R.,  Gorecki A.,  Marshall P.~J.,  Koopmans L.~V.,  Bolton
  A.~S.,  Moustakas L.~A.,   Burles S.,  2009, \mn@doi [Astrophysical Journal]
  {10.1088/0004-637X/690/1/670}, 690, 670

\bibitem[\protect\citeauthoryear{{Valageas}, {Barber}  \& {Munshi}}{{Valageas}
  et~al.}{2004}]{Valageas2004}
{Valageas} P.,  {Barber} A.~J.,   {Munshi} D.,  2004, \mn@doi [\mnras]
  {10.1111/j.1365-2966.2004.07248.x}, \href
  {https://ui.adsabs.harvard.edu/abs/2004MNRAS.347..654V} {347, 654}

\bibitem[\protect\citeauthoryear{{Valls-Gabaud}, {Cabanac}, {Lidman}, {Diego}
  \& {Saha}}{{Valls-Gabaud} et~al.}{2006}]{Valls2006}
{Valls-Gabaud} D.,  {Cabanac} R.,  {Lidman} C.,  {Diego} J.~M.,   {Saha} P.,
  2006, in {Mamon} G.~A.,  {Combes} F.,  {Deffayet} C.,   {Fort} B.,  eds,  EAS
  Publications Series Vol. 20, EAS Publications Series. pp 149--152,
  \mn@doi{10.1051/eas:2006062}

\bibitem[\protect\citeauthoryear{Van De~Vyvere, Sluse, Mukherjee, Xu  \&
  Birrer}{Van De~Vyvere et~al.}{2020}]{VanDeVyvere2020}
Van De~Vyvere L.,  Sluse D.,  Mukherjee S.,  Xu D.,   Birrer S.,  2020, \mn@doi
  [Astronomy and Astrophysics] {10.1051/0004-6361/202038942}, 644, 1

\bibitem[\protect\citeauthoryear{Van De~Vyvere, Gomer, Sluse, Xu, Birrer, Galan
   \& Vernardos}{Van De~Vyvere et~al.}{2022}]{VanDeVyvere2022}
Van De~Vyvere L.,  Gomer M.~R.,  Sluse D.,  Xu D.,  Birrer S.,  Galan A.,
  Vernardos G.,  2022, \mn@doi [Astronomy and Astrophysics]
  {10.1051/0004-6361/202141551}, 659, 1

\bibitem[\protect\citeauthoryear{Van~Rossum \& Drake}{Van~Rossum \&
  Drake}{2009}]{python}
Van~Rossum G.,  Drake F.~L.,  2009, Python 3 Reference Manual.
CreateSpace, Scotts Valley, CA

\bibitem[\protect\citeauthoryear{{Van de Vyvere}, {Sluse}, {Gomer}  \&
  {Mukherjee}}{{Van de Vyvere} et~al.}{2022}]{VanDeVyvere2022a}
{Van de Vyvere} L.,  {Sluse} D.,  {Gomer} M.~R.,   {Mukherjee} S.,  2022,
  \mn@doi [\aap] {10.1051/0004-6361/202243382}, \href
  {https://ui.adsabs.harvard.edu/abs/2022A&A...663A.179V} {663, A179}

\bibitem[\protect\citeauthoryear{Van~der Walt, Sch{\"o}nberger, Nunez-Iglesias,
  Boulogne, Warner, Yager, Gouillart  \& Yu}{Van~der Walt
  et~al.}{2014}]{scikit-image}
Van~der Walt S.,  Sch{\"o}nberger J.~L.,  Nunez-Iglesias J.,  Boulogne F.,
  Warner J.~D.,  Yager N.,  Gouillart E.,   Yu T.,  2014, PeerJ, 2, e453

\bibitem[\protect\citeauthoryear{Vegetti \& Koopmans}{Vegetti \&
  Koopmans}{2009}]{Vegetti2009}
Vegetti S.,  Koopmans L.~V.,  2009, \mn@doi [Monthly Notices of the Royal
  Astronomical Society] {10.1111/j.1365-2966.2009.15559.x}, 400, 1583

\bibitem[\protect\citeauthoryear{Vegetti, Koopmans, Bolton, Treu  \&
  Gavazzi}{Vegetti et~al.}{2010}]{Vegetti2010}
Vegetti S.,  Koopmans L.~V.,  Bolton A.,  Treu T.,   Gavazzi R.,  2010, \mn@doi
  [Monthly Notices of the Royal Astronomical Society]
  {10.1111/j.1365-2966.2010.16865.x}, 408, 1969

\bibitem[\protect\citeauthoryear{{Vegetti}, {Koopmans}, {Auger}, {Treu}  \&
  {Bolton}}{{Vegetti} et~al.}{2014}]{Vegetti2014b}
{Vegetti} S.,  {Koopmans} L.~V.~E.,  {Auger} M.~W.,  {Treu} T.,   {Bolton}
  A.~S.,  2014, \mn@doi [\mnras] {10.1093/mnras/stu943}, \href
  {https://ui.adsabs.harvard.edu/abs/2014MNRAS.442.2017V} {442, 2017}

\bibitem[\protect\citeauthoryear{{Virtanen} et~al.,}{{Virtanen}
  et~al.}{2020}]{scipy}
{Virtanen} P.,  et~al., 2020, \mn@doi [Nature Methods]
  {10.1038/s41592-019-0686-2}, \href {https://rdcu.be/b08Wh} {17, 261}

\bibitem[\protect\citeauthoryear{Witt \& Mao}{Witt \& Mao}{1997}]{Witt1997}
Witt H.~J.,  Mao S.,  1997, \mn@doi [Monthly Notices of the Royal Astronomical
  Society] {10.1093/mnras/291.1.211}, 291, 211

\bibitem[\protect\citeauthoryear{{Wong}, {Keeton}, {Williams}, {Momcheva}  \&
  {Zabludoff}}{{Wong} et~al.}{2011}]{Wong2011}
{Wong} K.~C.,  {Keeton} C.~R.,  {Williams} K.~A.,  {Momcheva} I.~G.,
  {Zabludoff} A.~I.,  2011, \mn@doi [\apj] {10.1088/0004-637X/726/2/84}, \href
  {https://ui.adsabs.harvard.edu/abs/2011ApJ...726...84W} {726, 84}

\bibitem[\protect\citeauthoryear{Wong et~al.,}{Wong et~al.}{2019}]{Wong2019}
Wong K.~C.,  et~al., 2019, \mn@doi [Monthly Notices of the Royal Astronomical
  Society] {10.1093/mnras/stz3094}, 498, 1420–1439

\bibitem[\protect\citeauthoryear{Xu, Sluse, Schneider, Springel, Vogelsberger,
  Nelson  \& Hernquist}{Xu et~al.}{2016}]{Xu2016}
Xu D.,  Sluse D.,  Schneider P.,  Springel V.,  Vogelsberger M.,  Nelson D.,
  Hernquist L.,  2016, \mn@doi [Monthly Notices of the Royal Astronomical
  Society] {10.1093/mnras/stv2708}, 456, 739

\bibitem[\protect\citeauthoryear{{Zhang}, {Lee}, {Krolewski}, {Shi}, {Horowitz}
   \& {Kooistra}}{{Zhang} et~al.}{2022}]{Zhang2022}
{Zhang} B.,  {Lee} K.-G.,  {Krolewski} A.,  {Shi} J.,  {Horowitz} B.,
  {Kooistra} R.,  2022, arXiv e-prints, \href
  {https://ui.adsabs.harvard.edu/abs/2022arXiv221109331Z} {p. arXiv:2211.09331}

\bibitem[\protect\citeauthoryear{Zhao}{Zhao}{1996}]{Zhao1996}
Zhao H.,  1996, \mn@doi [MNRAS] {10.1093/mnras/278.2.488}, 278, 488

\bibitem[\protect\citeauthoryear{{van der Walt}, {Colbert}  \&
  {Varoquaux}}{{van der Walt} et~al.}{2011}]{numpy}
{van der Walt} S.,  {Colbert} S.~C.,   {Varoquaux} G.,  2011, \mn@doi [Comput
  Sci Eng] {10.1109/MCSE.2011.37}, 13, 22

\makeatother
\end{thebibliography}

\appendix
\section{Systematic Tests of Shear Measurement}\label{AppA}

\begin{figure*}
    \centering
    \includegraphics[width=0.47\linewidth]{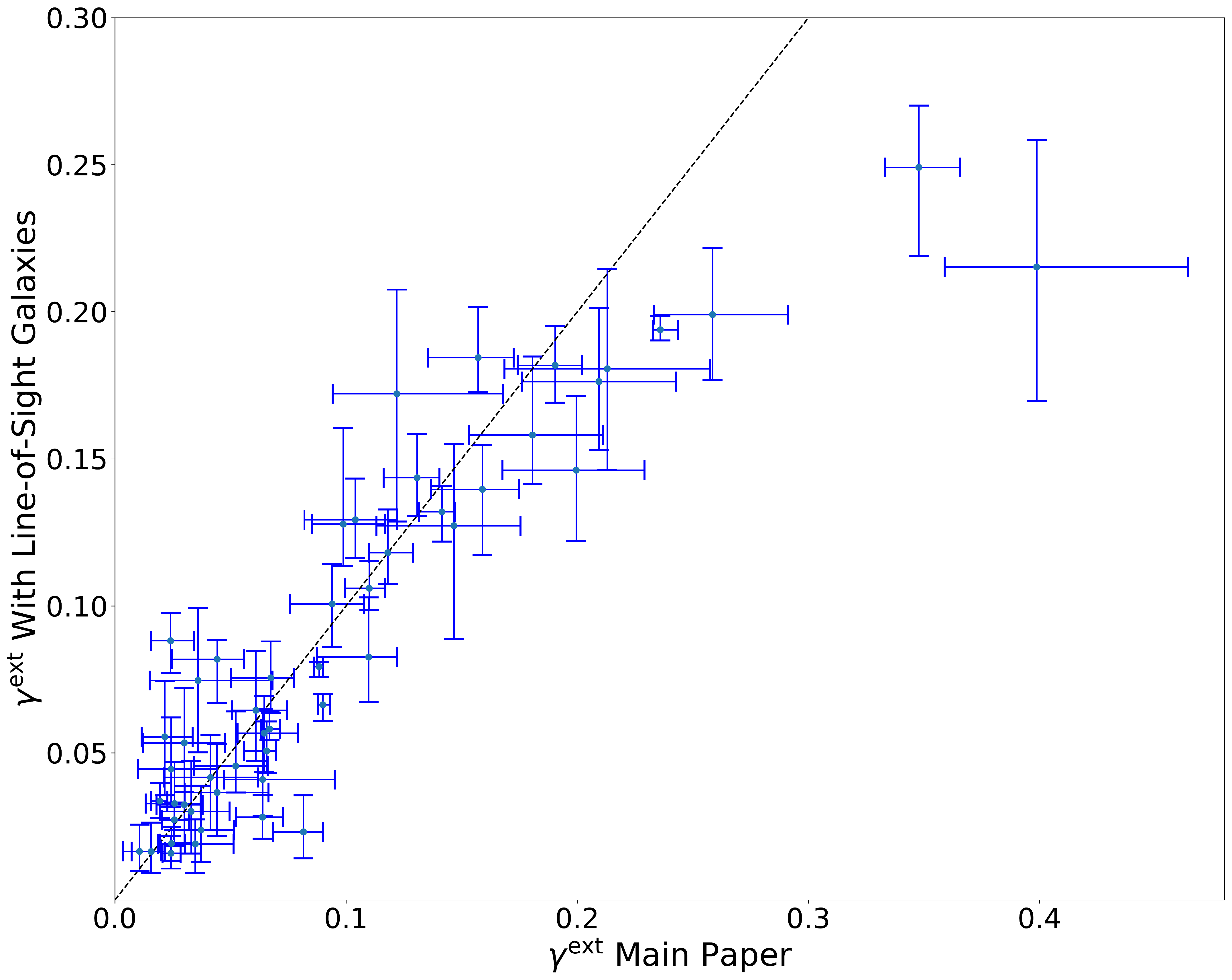}
    \includegraphics[width=0.47\linewidth]{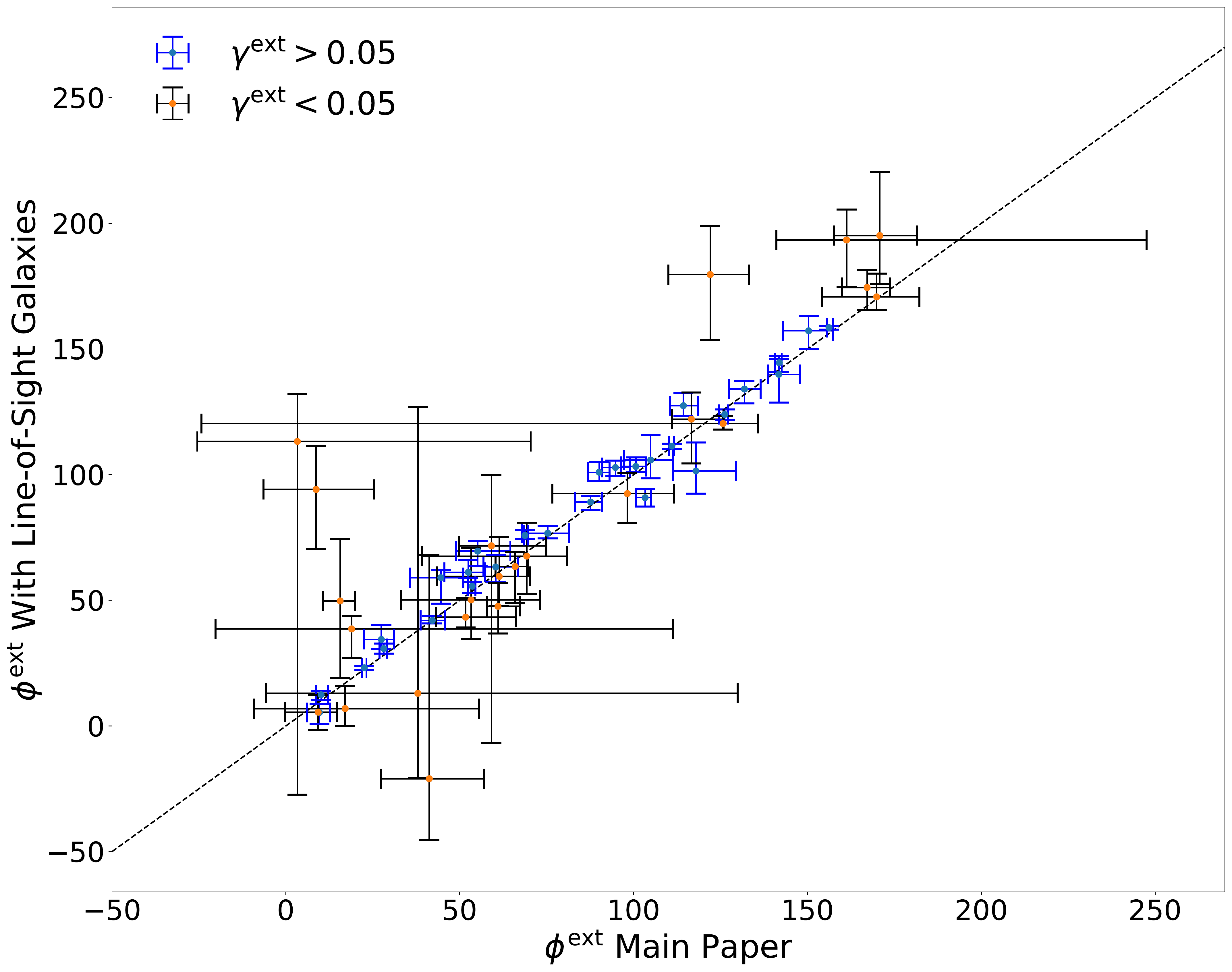}
    \caption{Best-fit external shear amplitude $\gamma^{\rm ext}$ and position angle $\phi^{\rm ext}$ for every lens in the sample. Shear values are shown for the lens models presented in the main paper which do not include nearby line-of-sight galaxies and for an additional set of lens model fitted, which include up to two nearby galaxies as SISs. Error bars are at 1$\sigma$ confidence intervals. The lens light subtraction and source analysis of every fit are the same, enabling a direct comparison of how the shear values change when nearby galaxies are included in the mass model. The inferred $\gamma^{\rm ext}$ and $\phi^{\rm ext}$ values are consistent with one another in the vast majority of lenses, irrespective of whether line-of-sight galaxies are included. In the right figure, models with $\gamma^{\rm ext} < 0.05$ are shown in black, illustrating that the measured values are only inconsistent for low shear lenses. Lensing due to nearby galaxies therefore is not responsible for the large shears which align with the lens mass model found in the main paper.}
    \label{Figure: appendix_a}
\end{figure*}

\begin{figure*}
     
    \centering
    \includegraphics[width=0.47\linewidth]{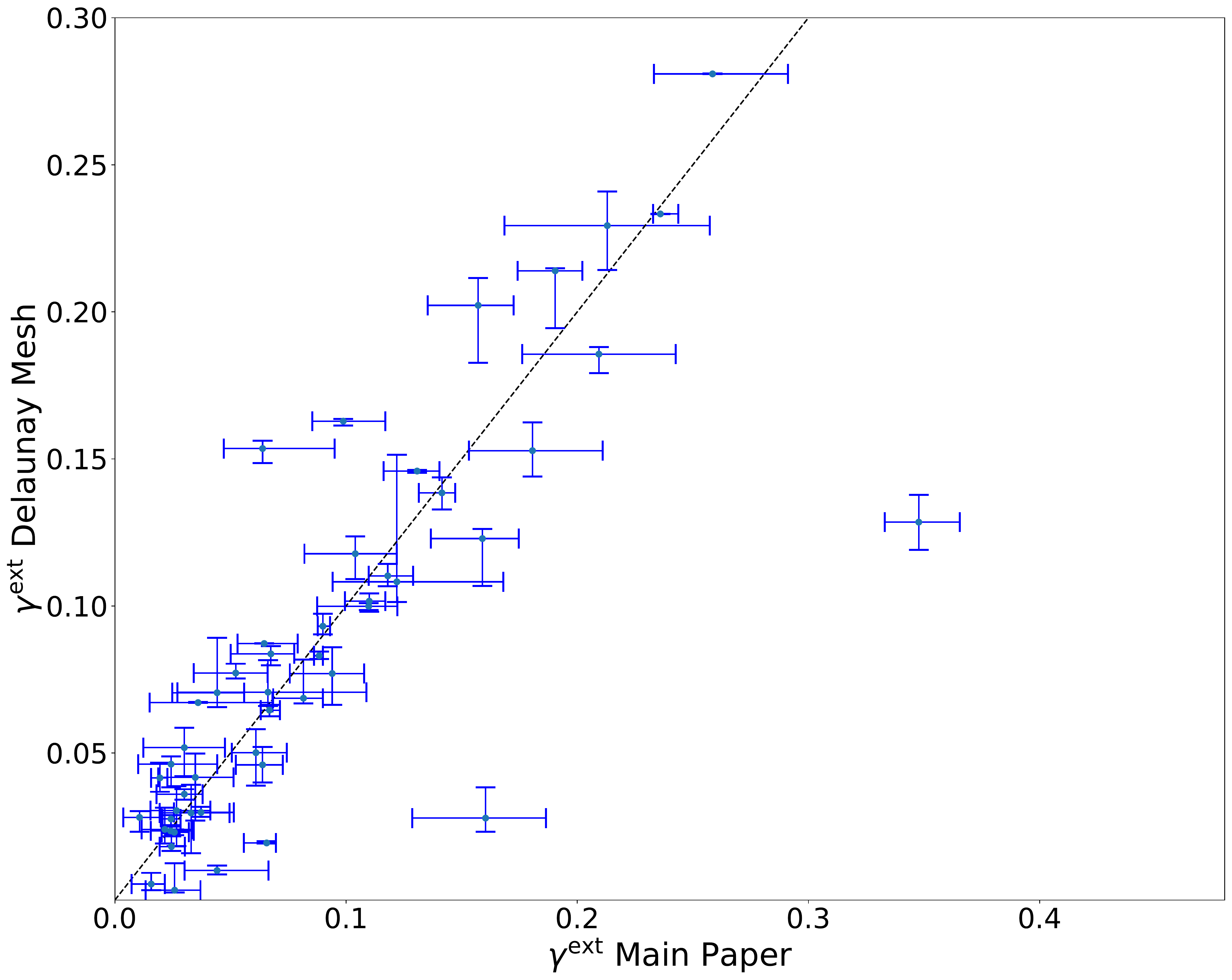}
    \includegraphics[width=0.47\linewidth]{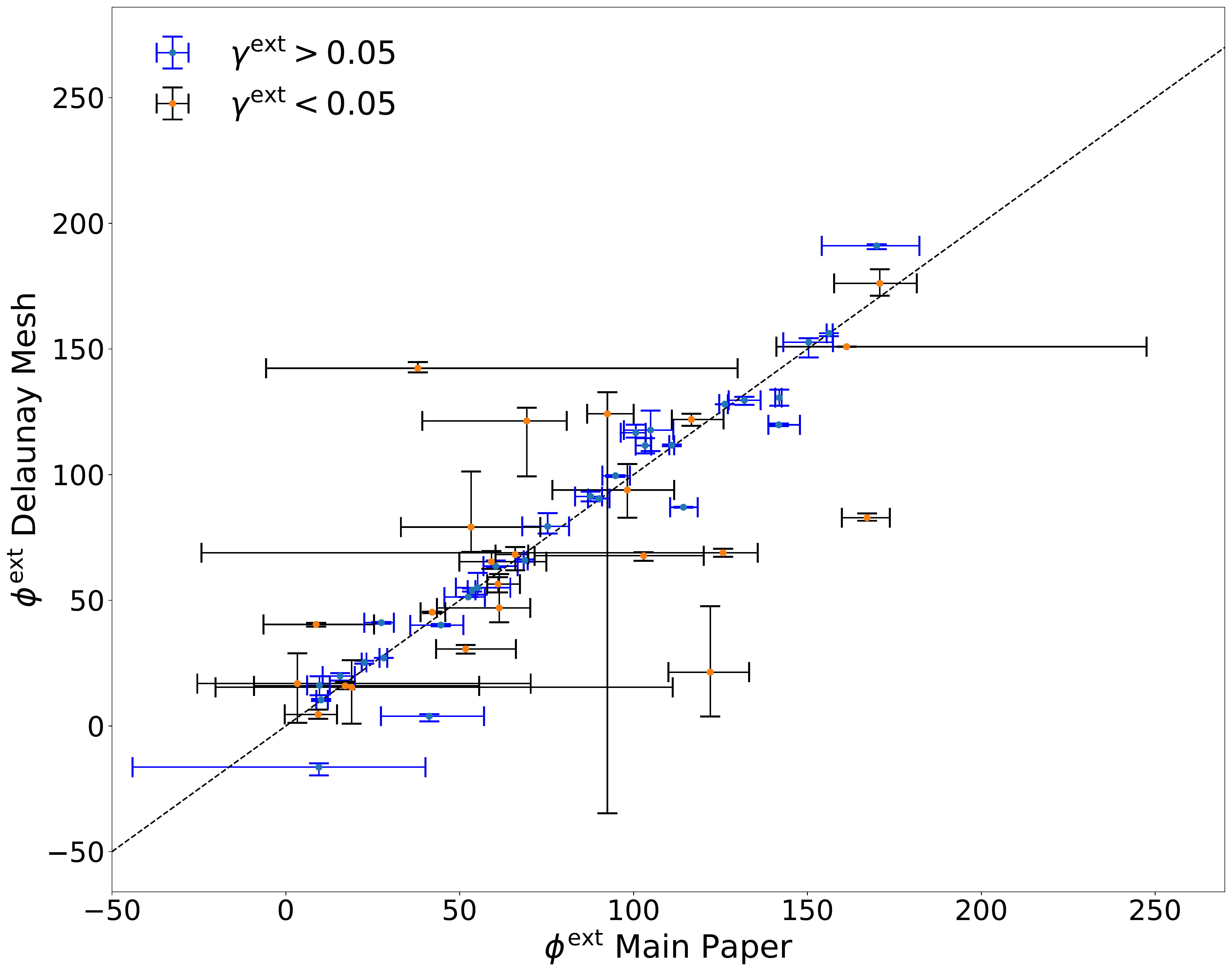}
    \caption{Best-fit external shear amplitude $\gamma^{\rm ext}$ and position angle $\phi^{\rm ext}$ for every lens in the sample. Shear values are shown for the lens models presented in the main paper which use a Voronoi mesh with luminosity based regularization and for an additional set of model fitted, which use a Delaunay mesh with interpolation regularization (see appendix A of \citealt{Nightingale2023}). Error bars are at 1$\sigma$ confidence intervals. In the right figure, models with $\gamma^{\rm ext} < 0.05$ are shown in black, illustrating that the measured values are only inconsistent for low shear lenses. The inferred $\gamma^{\rm ext}$ and $\phi^{\rm ext}$ values are consistent with one another in the vast majority of lenses, irrespective of the source model. The source analysis therefore does not impact our main conclusions.}
    \label{Figure: appendix_a2}
\end{figure*}

Real external shear is caused by galaxies or clusters of galaxies adjacent to the line of sight \citep{Wong2011}. It is possible that individual galaxies impart a localised external shear that is unmeasurable by weak lensing, which averages over large spatial scales. Here we repeat our analysis of every SLACS and GALLERY lens, but including adjacent galaxies in the mass model. Crucially, we use identical lens light subtraction and source pixellization, so we can directly compare the inferred shears.

We select which adjacent galaxies to include subjectively, based on their proximity and size. To implement this, we extend this study's $LP^{\rm 1}$ pipeline \citep{Etherington2022a} with a GUI where we look at $10\arcsec$ cut-outs of each lens and click on up to two  galaxies. These are added to the mass model as spherical isothermal spheres (SIS, an SIE with $q=1$), fixed to the centre of their brightest pixel and the same redshift as the main lens. Their Einstein radius $\theta^{\rm mass}_{\rm E}$ is given a flat prior from $0.0\arcsec$ to $0.5\arcsec$ and fitted with all other parameters. $\theta^{\rm mass}_{\rm E} = 0.5\arcsec$ corresponds to a mass far greater than that suggested by each galaxy's luminosity. It is an intentional choice not to use more informative priors, so that we can investigate how line-of-sight galaxies change the shear inference with maximal freedom. 

Adjacent galaxies do not significantly change the inferred external shear (Figure~\ref{Figure: appendix_a}). For almost every lens, the best-fit external shear for models with and without adjacent galaxies remains consistent. Large amplitudes of external shear are still preferred for many lenses, and their position angles still show the same alignment with the mass model position angles (not shown). The very largest external shears $\gamma_\mathrm{ext}\gtrsim0.3$ may be reduced, with a small statistical sample but with the two highest ($\sim$$0.40$ and $\sim$$0.36$) becoming $\sim$$0.21$ and $\sim$$0.24$ after accounting for adjacent galaxies. Such values remain physically infeasible. These results confirm that line-of-sight galaxies cannot explain the conclusions of this paper.

%Figure \ref{Figure: appendix_a} shows a comparison of the shear magnitudes inferred for the lens models presented in the main paper (which do not include line-of-sight galaxies) and the models which include line-of-sight galaxies. In nearly every lens, the shear magnitudes and angles are consistent with one another at one sigma confidence intervals. Large shear magnitudes are again seen in many lenses, and inspection of their position angles shows the same alignment with those of the mass model position angles (not shown). For the lenses with the two highest shear magnitudes ($\sim 0.4$ and $\sim 0.36$), including nearby galaxies reduces their shears ($\sim 0.21$ and $\sim 0.24$). These results confirm that line-of-sight galaxies cannot explain the results presented in the main paper.}

\section{Sensitivity to Source Model}

We investigate whether the shear measurements presented in the main paper depend on assumptions related to the source model. In the main paper, we followed \citet{Etherington2022a} in reconstructing the source using a Voronoi pixel mesh and a regularization scheme that adapted to the unlensed source luminosity \citep[see][]{Nightingale2018}. 
As an alternative, here we model the same lenses using a Delaunay pixel mesh and the interpolation-based regularization scheme described in appendix~A of \citet{Nightingale2022}.

The best-fit external shear with this alternative source-plane pixellization has amplitude and angle that is consistent with those in the main paper, for nearly every lens (see Figure~\ref{Figure: appendix_a2}). Our main conclusions are thus unaffected by even substantial changes in source reconstruction.

%Figure \ref{Figure: appendix_a2} shows a comparison of the shear magnitudes and angles inferred in the main paper and for fits using the Delaunay mesh and interpolation-based regularization. In nearly every lens, the shear magnitudes and angles are once again consistent at one sigma confidence. Assumptions regarding the source reconstruction therefore do not impact our main conclusions.}

\label{lastpage}

\end{document}